\documentclass[twocolumn,superscriptaddress,letter]{revtex4}%
\usepackage{amssymb}
\usepackage{color}
\usepackage{graphicx}
\usepackage{dcolumn}
\usepackage{bm}
\usepackage[header,title,page,titletoc]{appendix}
\usepackage{amsmath}
\usepackage{subfigure}
\usepackage{amsfonts}%
\providecommand{\U}[1]{\protect \rule{.1in}{.1in}}

\begin{document}
\title{Topological Interplay between Knots and Entangled Vortex-Membranes}
\author{Su-Peng Kou}
\thanks{Corresponding author}
\email{spkou@bnu.edu.cn}
\affiliation{Department of Physics, Beijing Normal University, Beijing, 100875, P. R. China}


\begin{abstract}
In this paper, the Kelvin wave and knot dynamics are studied on three
dimensional smoothly deformed entangled vortex-membranes in five dimensional
space. Owing to the existence of local Lorentz invariance and diffeomorphism
invariance, in continuum limit gravity becomes an emergent phenomenon on 3+1
dimensional zero-lattice (a lattice of projected zeroes): On the one hand, the
deformed zero-lattice can be denoted by curved space-time for knots; on the
other hand, the knots as topological defect of 3+1 dimensional zero-lattice
indicates matter may curve space-time. This work would help researchers to
understand the mystery in gravity.

\end{abstract}
\maketitle

\section{Introduction}

A vortex (point-vortex, vortex-line, vortex-membrane) consists of the rotating
motion of fluid around a common centerline. It is defined by the vorticity in
the fluid, which measures the rate of local fluid rotation. In three
dimensional (3D) superfluid (SF), the quantization of the vorticity manifests
itself in the quantized circulation $%
{\displaystyle \oint}
\mathbf{v}\cdot d\mathbf{l}=\frac{h}{m}$ where $h$ is Planck constant and $m$
is atom mass of SF. Vortex-lines can twist around its equilibrium position
(common centerline) forming a transverse and circularly polarized wave (Kelvin
wave)\cite{Thomson1880a,Donnelly1991a}. Because Kelvin waves are relevant to
Kolmogorov-like turbulence\cite{S95,V00}, a variety of approaches have been
used to study this phenomenon. For two vortex-lines, owing to the interaction,
the leapfrogging motion has been predicted in classical fluids from the works
of Helmholtz and Kelvin\cite{dys93, hic22, bor13,wac14, cap14,1}. Another
interesting issue is entanglement between two vortex-lines. In mathematics,
vortex-line-entanglement can be characterized by knots with different linking
numbers. The study of knotted vortex-lines and their dynamics has attracted
scientists from diverse settings, including classical fluid dynamics and
superfluid dynamics\cite{Kleckner2013,Hall2016}.

In the paper\cite{kou}, the Kelvin wave and knot dynamics in high dimensional
vortex-membranes were studied, including the leapfrogging motion and the
entanglement between two vortex-membranes. A new theory - \emph{knot physics}
is developed to characterize the entanglement evolution of 3D leapfrogging
vortex-membranes in five-dimensional (5D) inviscid incompressible fluid.
According to knot physics, it is the 3D quantum Dirac model that describes the
knot dynamics of leapfrogging vortex-membranes (we have called it
knot-crystal, that is really plane Kelvin-waves with fixed wave-length). The
knot physics may give a complete interpretation on quantum mechanics.

In this paper, we will study the Kelvin wave and knot dynamics on 3D deformed
knot-crystal, particularly the topological interplay between knots and the
lattice of projected zeroes (we call it zero-lattice). Owing to the existence
of local Lorentz invariance and diffeomorphism invariance, the gravitational
interaction emerges: On the one hand, the deformed zero-lattice can be denoted
by curved space-time; on the other hand, the knots deform the zero-lattice
that indicates matter may curve space-time (see below discussion).

The paper is organized as below. In Sec. II, we introduce the concept of
"zero-lattice" from projecting a knot-crystal. In addition, to characterize
the entangled vortex-membranes, we introduce geometric space and winding
space. In Sec. III, we derive the massive Dirac model in the
vortex-representation of knot states on geomatric space and that on winding
space. In Sec. IV, we consider the deformed knot-crystal as a background and
map the problem onto Dirac fermions on a curved space-time. In Sec. V, the
gravity in knot physics emerges as a topological interplay between
zero-lattice and knots and the knot dynamics on deformed knot-crystal is
described by Einstein's general relativity. Finally, the conclusions are drawn
in Sec. VI.

\section{Knot-crystal and the corresponding zero-lattice}

\subsection{Knot-crystal}

Knot-crystal is a system of two periodically entangled vortex-membranes that
is described by a special pure state of Kelvin waves with fixed wave length
$\mathbf{Z}_{\mathrm{knot-crystal}}(\vec{x},t)$\cite{kou}. In emergent quantum
mechanics, we consider knot-crystal as a ground state for excited knot states,
i.e.,
\begin{equation}
\mathbf{Z}_{\mathrm{knot-crystal}}(\vec{x},t)=\left(
\begin{array}
[c]{c}%
\mathrm{z}_{\mathrm{A}}(\vec{x},t)\\
\mathrm{z}_{\mathrm{B}}(\vec{x},t)
\end{array}
\right)  \rightarrow \left \vert \text{\textrm{vacuum}}\right \rangle .
\end{equation}
On the one hand, a knot is a piece of knot-crystal and becomes a topological
excitation on it; On the other hand, a knot-crystal can be regarded as a
composite system with multi-knot, each of which is described by same tensor state.

Because a knot-crystal is a plane Kelvin wave with fixed wave vector $k_{0}$,
we can use the tensor representation to characterize knot-crystals\cite{kou},
\begin{equation}
\mathbf{\vec{\Gamma}}_{\mathrm{knot-crystal}}^{I}=(\vec{n}_{\sigma}%
^{I}\mathbf{\sigma}^{I})\otimes(\vec{n}_{\tau}\mathbf{\tau}+\vec{1}\tau_{0})
\end{equation}
where $\vec{1}=\left(
\begin{array}
[c]{cc}%
1 & 0\\
0 & 1
\end{array}
\right)  $ and $\mathbf{\sigma}^{I}$, $\mathbf{\tau}^{I}$ are $2\times2$ Pauli
matrices for helical and vortex degrees of freedom, respectively. For example,
a particular knot-crystal is called SOC knot-crystal $\mathbf{Z}%
_{\mathrm{knot-crystal}}(\vec{x})$\cite{kou}, of which the tensor state is
given by
\begin{align}
\left \langle \mathbf{\sigma}^{X}\otimes \vec{1}\right \rangle  &  =\vec
{n}_{\sigma}^{X}=(1,0,0),\\
\left \langle \mathbf{\sigma}^{Y}\otimes \vec{1}\right \rangle  &  =\vec
{n}_{\sigma}^{Y}=(0,1,0),\nonumber \\
\left \langle \mathbf{\sigma}^{Z}\otimes \vec{1}\right \rangle  &  =\vec
{n}_{\sigma}^{Z}=(0,0,1).\nonumber
\end{align}
For the SOC knot-crystal, along x-direction, the plane Kelvin wave becomes
$\mathrm{z}(x)=\sqrt{2}r_{0}\cos(k_{0}\cdot x);$ along y-direction, the plane
Kelvin wave becomes $\mathrm{z}(y)=\frac{1}{\sqrt{2}}r_{0}(e^{ik\cdot
y}+ie^{-ik\cdot y});$ along z-direction, the plane Kelvin wave becomes
$\mathrm{z}(z)=r_{0}e^{ik\cdot z}.$

For a knot-crystal, another important property is generalized spatial
translation symmetry that is defined by the translation operation
$\mathcal{T}(\Delta x^{I})=e^{i\cdot(\hat{k}_{0}^{I}\cdot \Delta x^{I}%
)\cdot \mathbf{\vec{\Gamma}}_{\mathrm{knot-crystal}}^{I}}$
\begin{align}
\mathbf{Z}(x^{I},t)  &  \rightarrow \mathcal{T}(\Delta x^{I})\mathbf{Z}%
(x^{i},t)\nonumber \\
&  =e^{i\cdot(\hat{k}_{0}^{I}\cdot \Delta x^{i})\cdot \mathbf{\vec{\Gamma}%
}_{\mathrm{knot-crystal}}^{I}}\mathbf{Z}(x^{i},t).
\end{align}
Here $\hat{k}^{I}$ is $-i\frac{d}{dx^{I}}$ ($I=x,y,z$). For example, for the
knot states on 3D SOC knot-crystal, the translation operation along $x^{I}%
$-direction becomes
\begin{equation}
\mathcal{T}(\Delta x^{I})=e^{i(\hat{k}^{I}\cdot \Delta x^{I})\cdot(\sigma
^{I}\otimes \vec{1})}.
\end{equation}

\subsection{Winding space and geometric space}

For a knot-crystal, we can study it properties on a 3D space ($x,y,z$). In the
following part, we call the space of ($x,y,z$) \emph{geometric space}.
According to the generalized spatial translation symmetry, each spatial point
($x,y,z$) in geometric space corresponds to a point denoted by three winding
angles $(\Phi_{x}(x),\Phi_{y}(y),\Phi_{z}(z))$ where $\Phi_{x^{I}}(x^{I})$ is
the winding angle along $x^{I}$-direction. As a result, we may use the winding
angles along different directions to denote a given point $\vec{\Phi}(\vec
{x})=(\Phi_{x}(x),\Phi_{y}(y),\Phi_{z}(z)).$ We call the space of winding
angles $(\Phi_{x}(x),\Phi_{y}(y),\Phi_{z}(z))$ \emph{winding space}. See the
illustration in Fig.1(d).

\begin{figure}[ptb]
\includegraphics[clip,width=0.53\textwidth]{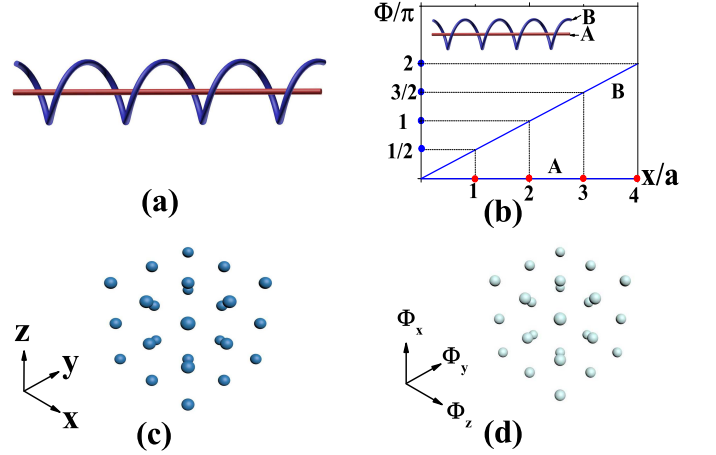}\caption{(a) An
illustration of a 1D knot-crystal; (b) The relationship between winding angle
$\Phi$ and coordinate position $x$. The red dots consist of a 1D zero-lattice
in geometric space and the blue dots consist of a zero-lattice in winding
space; (c) An illustration of a 3D uniform zero-lattice in geometric space;
(d) An illustration of a 3D uniform zero-lattice in winding space.}%
\end{figure}

For a 1D leapfrogging knot-crystal that describes two entangled vortex-lines
with leapfrogging motion, the function is given by
\begin{equation}
\mathbf{Z(}\vec{x},t\mathbf{)}=r_{0}\left(
\begin{array}
[c]{c}%
\cos(\frac{\omega^{\ast}t}{2})\\
-i\sin(\frac{\omega^{\ast}t}{2})
\end{array}
\right)  e^{i\frac{\pi}{a}x}e^{-i\omega_{0}t+i\omega^{\ast}t/2},
\end{equation}
where $\omega^{\ast}$ is angular frequency of leapfrogging motion. For the 1D
$\sigma_{z}$-knot-crystal, the coordinate on winding space is $\Phi
(x)=\frac{\pi}{a}x.$ Another example is 3D SOC knot-crystal\cite{1}, of which
the function is given by
\begin{align}
\mathbf{Z}_{\mathrm{KC}}\mathbf{(}\vec{x},t\mathbf{)}  &  =\left(
\begin{array}
[c]{c}%
\mathrm{z}_{\mathrm{KC,A}}(\vec{x},t)\\
\mathrm{z}_{\mathrm{KC,B}}(\vec{x},t)
\end{array}
\right)  =r_{0}\left(
\begin{array}
[c]{c}%
\cos(\frac{\omega^{\ast}t}{2})\\
-i\sin(\frac{\omega^{\ast}t}{2})
\end{array}
\right)  e^{-i\omega_{0}t+i\omega^{\ast}t/2}\nonumber \\
&  \cdot \sqrt{2}r_{0}\cos(\Phi_{x}(x))\cdot(\frac{1}{\sqrt{2}}r_{0}%
(e^{i\Phi_{y}(y)}+ie^{-i\Phi_{y}(y)}))e^{i\Phi_{z}(z)},
\end{align}
where the coordinates on winding space are $\Phi_{x}(x)=\frac{\pi}{a}x,$
$\Phi_{y}(y)=\frac{\pi}{a}y,$ $\Phi_{z}(z)=\frac{\pi}{a}z,$ respectively.

In addition, there exists generalized spatial translation symmetry on winding
space. On winding space, the translation operation $\mathcal{T}(\Delta \Phi
^{I})$ becomes
\begin{equation}
\mathcal{T}(\Delta \Phi^{I})=e^{i\cdot \sum_{I}\Delta \Phi^{I}\cdot
\mathbf{\vec{\Gamma}}_{\mathrm{knot-crystal}}^{I}}\nonumber
\end{equation}
where $\Delta \Phi^{I}$ denotes the distance on winding space.

\subsection{Zero-lattice}

Before introduce zero-lattice, we firstly review the projection between two
entangled vortex-membranes $\mathrm{z}_{\mathrm{A/B}}(\vec{x},t)=\xi
_{\mathrm{A/B}}(\vec{x},t)+i\eta_{\mathrm{A/B}}(\vec{x},t)$ along a given
direction $\theta$ in 5D space by
\begin{equation}
\hat{P}_{\theta}\left(
\begin{array}
[c]{c}%
\xi_{\mathrm{A/B}}(\vec{x},t)\\
\eta_{\mathrm{A/B}}(\vec{x},t)
\end{array}
\right)  =\left(
\begin{array}
[c]{c}%
\xi_{\mathrm{A/B},\theta}(\vec{x},t)\\
\left[  \eta_{\mathrm{A/B},\theta}(\vec{x},t)\right]  _{0}%
\end{array}
\right)
\end{equation}
where $\xi_{\mathrm{A/B},\theta}(\vec{x},t)=\xi_{\mathrm{A/B}}(\vec{x}%
,t)\cos \theta+\eta_{\mathrm{A/B}}(\vec{x},t)\sin \theta$ is variable and
$\left[  \eta_{\mathrm{A/B},\theta}(\vec{x},t)\right]  _{0}=\xi_{\mathrm{A/B}%
}(\vec{x},t)\sin \theta-\eta_{\mathrm{A/B}}(\vec{x},t)\cos \theta$ is constant.
So the projected vortex-membrane is described by the function $\xi
_{\mathrm{A/B},\theta}(\vec{x},t).$ For two projected vortex-membranes
described by $\xi_{\mathrm{A},\theta}(\vec{x},t)$ and $\xi_{\mathrm{B},\theta
}(\vec{x},t),$ a zero is solution of the equation
\begin{align}
\hat{P}_{\theta}[\mathrm{z}_{\mathrm{A}}(\vec{x},t)]  &  \equiv \xi
_{\mathrm{A},\theta}(\vec{x},t)\\
&  =\hat{P}_{\theta}[\mathrm{z}_{\mathrm{B}}(\vec{x},t)]\equiv \xi
_{\mathrm{B},\theta}(\vec{x},t).\nonumber
\end{align}

After projection, the knot-crystal becomes a zero lattice. For example, a 1D
leapfrogging\ knot-crystal is described by
\begin{equation}
\mathbf{Z}_{\mathrm{KC}}\mathbf{(}\vec{x},t\mathbf{)}=r_{0}\left(
\begin{array}
[c]{c}%
\cos(\frac{\omega^{\ast}t}{2})\\
-i\sin(\frac{\omega^{\ast}t}{2})
\end{array}
\right)  e^{i\frac{\pi}{a}x}e^{-i\omega_{0}t+i\omega^{\ast}t/2}.
\end{equation}
According to the knot-equation $\hat{P}_{\theta}[z_{\mathrm{KC,A}}(x)]=\hat
{P}_{\theta}[z_{\mathrm{KC,B}}(x)],$ we have
\begin{equation}
\bar{x}_{0}=a\cdot X+\frac{a}{\pi}\omega_{0}t
\end{equation}
where $\theta=-\frac{\pi}{2}$ and $\bar{x}_{0}$ is the position of zero. As a
result, we have a periodic distribution of zeroes (knots).

For a 3D leapfrogging SOC\ knot-crystal described by $\mathbf{Z}_{\mathrm{KC}%
}\mathbf{(}\vec{x},t\mathbf{)}=\left(
\begin{array}
[c]{c}%
\mathrm{z}_{\mathrm{KC,A}}(\vec{x},t)\\
\mathrm{z}_{\mathrm{KC,B}}(\vec{x},t)
\end{array}
\right)  ,$ we have similar situation -- the solution of zeroes doesn't change
when the tensor order changes, i.e., $\left \langle \mathbf{\sigma}\otimes
\vec{1}\right \rangle =\vec{n}_{\sigma}=(0,0,1)\rightarrow \vec{n}_{\sigma
}=(n_{x},n_{y},n_{x})$ with $\left \vert \vec{n}_{\sigma}\right \vert
=1$\cite{kou}. We call the periodic distribution of zeroes to be
\emph{zero-lattice}. See the illustration of a 1D zero-lattice in Fig.1(b) and
3D zero-lattice in Fig.1(c).

Along a given direction $\vec{e}$, after shifting the distance $a,$ the phase
angle of vortex-membranes in knot-crystal changes $\pi,$ i.e.,
\begin{equation}
\vec{\Phi}(\vec{x},t)\rightarrow \vec{\Phi}(\vec{x}+a\cdot \vec{e},t)=\vec{\Phi
}(\vec{x},t)+\pi.
\end{equation}
Thus, on the winding space, we have a corresponding "zero-lattice" of discrete
lattice sites described by the three integer numbers
\begin{equation}
\vec{X}=(X,Y,Z)=\frac{1}{\pi}\vec{\Phi}-\frac{1}{\pi}\vec{\Phi}%
\operatorname{mod}\pi.
\end{equation}
See the illustration of a 1D zero-lattice in Fig.1(b) and 3D zero-lattice in Fig.1(d).

\section{Dirac model for knot on zero-lattice}

\subsection{Dirac model on geometric space}

\subsubsection{Dirac model in sublattice-representation on geometric space}

It was known that in emergent quantum mechanics, a 3D SOC knot-crystal becomes
multi-knot system, of which the effective theory becomes a Dirac model in
quantum field theory. In emergent quantum mechanics, the Hamiltonian for a 3D
SOC knot-crystal has two terms -- the kinetic term from global winding and the
mass term from leapfrogging motion. Based on a representation of projected
state, a 3D SOC knot-crystal is reduced into a "two-sublattice" model with
discrete spatial translation symmetry, of which the knot states are described
by $\left \vert \mathrm{L}\right \rangle $ and $\left \vert \text{\textrm{R}%
}\right \rangle $ (or the Wannier states $c_{\mathrm{L},i}^{\dagger}\left \vert
\text{\textrm{vacuum}}\right \rangle $ and $c_{\mathrm{R},j}^{\dagger
}\left \vert \text{\textrm{vacuum}}\right \rangle $). We call it the Dirac model
in \emph{sublattice-representation}.

In sublattice-representation on geometric space, the equation of motion of
knots is determined by the Schr\"{o}dinger equation with the Hamiltonian
\begin{align}
\mathcal{H}_{\mathrm{knot}}  &  =\int(\psi^{\dagger}\mathrm{\hat{H}%
}_{\mathrm{knot}}\psi)d^{3}x,\nonumber \\
\mathrm{\hat{H}}_{\mathrm{knot}}  &  =-c_{\mathrm{eff}}\vec{\Gamma}\cdot
\vec{p}_{\mathrm{knot}}+m_{\mathrm{knot}}c_{\mathrm{eff}}^{2}\Gamma^{5},
\end{align}
where $\psi^{\dagger}(t,\vec{x})$ is an four-component fermion field as
$\psi^{\dagger}(t,\vec{x})=\left(
\begin{array}
[c]{cccc}%
\psi_{\uparrow L}^{\dagger}(t,\vec{x}) & \psi_{\uparrow R}^{\dagger}(t,\vec
{x}) & \psi_{\downarrow L}^{\dagger}(t,\vec{x}) & \psi_{\downarrow R}%
^{\dagger}(t,\vec{x})
\end{array}
\right)  $. Here, $L,R$ label two chiral-degrees of freedom that denote the
two possible sub-lattices, $\uparrow,\downarrow$ label two spin degrees of
freedom that denote the two possible winding directions. We have
\begin{equation}
\Gamma^{5}=\vec{1}\otimes \iota_{x}\mathbf{,}%
\end{equation}
and
\begin{align}
\Gamma^{1}  &  =\sigma^{x}\otimes \iota_{y},\text{ }\\
\Gamma^{2}  &  =\sigma^{y}\otimes \iota_{y},\nonumber \\
\Gamma^{3}  &  =\sigma^{z}\otimes \iota_{y}.\nonumber
\end{align}
$\vec{p}_{\mathrm{knot}}=\hbar_{\mathrm{knot}}\vec{k}$ is the momentum
operator. $m_{\mathrm{knot}}c_{\mathrm{eff}}^{2}=2\hbar_{\mathrm{knot}}%
\omega^{\ast}$ plays role of the mass of knots and $c_{\mathrm{eff}}%
=\frac{a\cdot J}{\hbar_{\mathrm{knot}}}=2a\omega_{0}$ play the role of light
speed where $a$ is a fixed length that denotes the half pitch of the windings
on the knot-crystal.

In addition, the low energy effective Lagrangian of knots on 3D SOC
knot-crystal is obtained as
\begin{equation}
\mathcal{L}_{\mathrm{3D}}=\bar{\psi}(i\gamma^{\mu}\hat{\partial}_{\mu
}-m_{\mathrm{knot}})\psi
\end{equation}
where $\bar{\psi}=\psi^{\dagger}\gamma^{0},$ $\gamma^{\mu}$ are the reduced
Gamma matrices,
\begin{equation}
\gamma^{1}=\gamma^{0}\Gamma^{1},\text{ }\gamma^{2}=\gamma^{0}\Gamma^{2},\text{
}\gamma^{3}=\gamma^{0}\Gamma^{3},
\end{equation}
and
\begin{equation}
\gamma^{0}=\Gamma^{5},\text{ }\gamma^{5}=i\gamma^{0}\gamma^{1}\gamma^{2}%
\gamma^{3}.
\end{equation}

\subsubsection{Dirac model in vortex-representation on geometric space}

\begin{figure}[ptb]
\includegraphics[clip,width=0.53\textwidth]{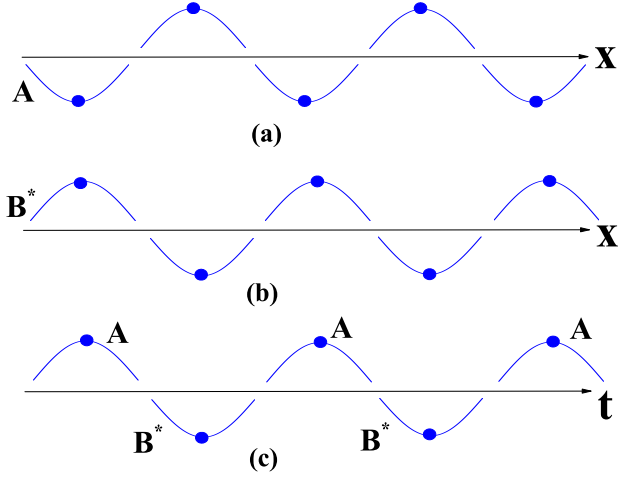}\caption{An illustration
of knot states in vortex-representation: $\mathrm{A}$ and $\mathrm{B}$ denote
two 1D vortex-lines. Here \textrm{B}$^{\ast}$ denotes conjugate representation
of vortex-line-\textrm{B}. The curves with blue dots denote knots on the
knot-crystal -- the curves with blue dot above the line are denoted by
$c_{i}^{\dagger}\left \vert 0\right \rangle $ and the curves with blue dot below
the line are denoted by $(c_{i}^{\dagger}\left \vert 0\right \rangle )^{\dagger
}$.}%
\end{figure}

In this paper, we derive the effective Dirac model for a knot-crystal based on
a representation of vortex degrees of freedom. We call it
\emph{vortex-representation}.

In Ref.\cite{kou}, it was known that a knot has four degrees of freedom, two
spin degrees of freedom $\uparrow$ or $\downarrow$ from the helicity degrees
of freedom, the other two vortex degrees of freedom from the vortex degrees of
freedom that characterize the vortex-membranes, $\mathrm{A}$ or $\mathrm{B}$.
The basis to define the microscopic structure of a knot is given by
$\left \vert \uparrow,\mathrm{A}\right \rangle ,$ $\left \vert \uparrow
,\mathrm{B}\right \rangle ,$ $\left \vert \downarrow,\mathrm{A}\right \rangle ,$
$\left \vert \downarrow,\mathrm{B}\right \rangle .$

We define operator of knot states by the region of the phase angle of a knot:
for the case of $\phi_{0}\operatorname{mod}(2\pi)\in(-\pi,0],$ we have
$c^{\dagger}\left \vert 0\right \rangle $; For the case of $\phi_{0}%
\operatorname{mod}(2\pi)\in(0,\pi],$ we have $(c^{\dagger}\left \vert
0\right \rangle )^{\dagger}$. As shown in Fig.2, we label the knots by Wannier
state $\left \vert i,\mathrm{A,\uparrow}\right \rangle ,$ $\left \vert
i+1,\mathrm{A,\uparrow}\right \rangle ^{\ast},$ $\left \vert
i+2,\mathrm{A,\uparrow}\right \rangle $, $\left \vert i+3,\mathrm{A,\uparrow
}\right \rangle ^{\ast}$...

To characterize the energy cost from global winding, we use an effective
Hamiltonian to describe the coupling between 2-knot states along $x^{I}%
$-direction on 3D SOC knot-crystal
\begin{equation}
Jc_{\mathrm{A/B}i}^{\dagger}T_{\mathrm{A/B,A/B}}^{I}c_{\mathrm{A/B},i+e^{I}}%
\end{equation}
with the annihilation operator of knots at the site $i,$ $c_{\mathrm{A/B}%
,i}=\left(
\begin{array}
[c]{c}%
c_{\mathrm{A/B,}\uparrow,i}\\
c_{\mathrm{A/B,}\downarrow,i}%
\end{array}
\right)  $. $J$ is the coupling constant between two nearest-neighbor knots.
According to the generalized translation symmetry, the transfer matrices
$T_{\mathrm{A/B,A/B}}^{I}$ along $x^{I}$-direction are defined by
\begin{equation}
T_{\mathrm{A,A}}^{I}=T_{\mathrm{B,B}}^{I}=e^{ia(\hat{k}^{I}\cdot \sigma^{I})}%
\end{equation}
and
\begin{equation}
T_{\mathrm{A,B}}^{I}=T_{\mathrm{B,A}}^{I}=0.
\end{equation}

After considering the spin rotation symmetry and the symmetry of
vortex-membrane-\textrm{A} and vortex-membrane-\textrm{B}, the effective
Hamiltonian from global winding energy can be described by a familiar
formulation%
\begin{equation}
\mathcal{H}_{\mathrm{coupling}}=\mathcal{\hat{H}}_{\mathrm{coupling,B}%
}+\mathcal{\hat{H}}_{\mathrm{coupling,A}}%
\end{equation}
where
\begin{equation}
\mathcal{\hat{H}}_{\mathrm{coupling,A}}=J%
{\displaystyle \sum \limits_{i,I}}
c_{\mathrm{A,}i}^{\dagger}e^{ia(\hat{k}^{I}\cdot \sigma^{I})}c_{\mathrm{A}%
,i+e^{I}}+h.c.
\end{equation}
and%
\begin{equation}
\mathcal{\hat{H}}_{\mathrm{coupling,B}}=J%
{\displaystyle \sum \limits_{i,I}}
c_{\mathrm{B,}i}^{\dagger}e^{ia(\hat{k}^{I}\cdot \sigma^{I})}c_{\mathrm{B}%
,i+e^{I}}+h.c.
\end{equation}
We then use path-integral formulation to characterize the effective
Hamiltonian for a knot-crystal as
\begin{equation}
\int \mathcal{D}\psi^{\dagger}(t,\vec{x})\mathcal{D}\psi(t)e^{i\mathcal{S}%
/\hbar}%
\end{equation}
where $\mathcal{S}=\int \mathcal{L}dt$ and $\mathcal{L}=i%
{\displaystyle \sum \limits_{i}}
\psi_{i}^{\dagger}\partial_{t}\psi_{i}-\mathcal{H}_{\mathrm{coupling}}$. To
describe the knot states on 3D knot-crystal, we have introduced a
four-component fermion field to be
\begin{equation}
\psi(\mathbf{x})=\left(
\begin{array}
[c]{c}%
\psi_{\mathrm{A,}\uparrow}(t,\vec{x})\\
\psi_{\mathrm{B,}\uparrow}(t,\vec{x})\\
\psi_{\mathrm{A,}\downarrow}(t,\vec{x})\\
\psi_{\mathrm{B,}\downarrow}(t,\vec{x})
\end{array}
\right)
\end{equation}
where $\mathrm{A},\mathrm{B}$ label vortex degrees of freedom and
$\uparrow,\downarrow$ label two spin degrees of freedom that denote the two
possible winding directions along a given direction $\vec{e}$.

In continuum limit, we have%
\begin{align}
\mathcal{H}_{\mathrm{coupling}}  &  =\mathcal{\hat{H}}_{\mathrm{coupling,B}%
}+\mathcal{\hat{H}}_{\mathrm{coupling,A}}\nonumber \\
&  =2aJ%
{\displaystyle \sum \limits_{k}}
\psi_{\mathrm{A,}k}^{\dagger}[\sigma_{x}\cos k_{x}+\sigma_{y}\cos k_{y}%
+\sigma_{z}\cos k_{z}]\psi_{\mathrm{A,}k}\nonumber \\
&  +2aJ%
{\displaystyle \sum \limits_{k}}
\psi_{\mathrm{B,}k}^{\dagger}[\sigma_{x}\cos k_{x}+\sigma_{y}\cos k_{y}%
+\sigma_{z}\cos k_{z}]\psi_{\mathrm{B,}k}%
\end{align}
where the dispersion of knots is
\begin{equation}
E_{\mathrm{A/B,}k}\simeq c_{\mathrm{eff}}[(\vec{k}-\vec{k}_{0})\cdot
\vec{\sigma}],
\end{equation}
where $\vec{k}_{0}=(\frac{\pi}{2},\frac{\pi}{2},\frac{\pi}{2})$ and
$c_{\mathrm{eff}}=2aJ$ is the velocity. In the following part we ignore
$\vec{k}_{0}$.

Next, we consider the mass term from leapfrogging motion, of which the angular
frequency $\omega^{\ast}$. For leapfrogging motion obtained by\cite{1}, the
function of the two entangled vortex-membranes at a given point in geometric
space is simplified by
\begin{equation}
\left(
\begin{array}
[c]{c}%
\mathrm{z}_{\mathrm{A}}(\vec{x}=0,t)\\
\mathrm{z}_{\mathrm{B}}(\vec{x}=0,t)
\end{array}
\right)  =\frac{r_{0}}{2}\left(
\begin{array}
[c]{c}%
1+e^{i\omega^{\ast}t}\\
1-e^{i\omega^{\ast}t}%
\end{array}
\right)  .
\end{equation}
At $t=0$, we have $\left(
\begin{array}
[c]{c}%
\mathrm{z}_{\mathrm{A}}(\vec{x},t)\\
\mathrm{z}_{\mathrm{B}}(\vec{x},t)
\end{array}
\right)  =\left(
\begin{array}
[c]{c}%
1\\
0
\end{array}
\right)  $; At $t=\frac{\pi}{\omega^{\ast}}$, we have $\left(
\begin{array}
[c]{c}%
\mathrm{z}_{\mathrm{A}}(\vec{x},t)\\
\mathrm{z}_{\mathrm{B}}(\vec{x},t)
\end{array}
\right)  =\left(
\begin{array}
[c]{c}%
0\\
1
\end{array}
\right)  .$ The leapfrogging knot-crystal leads to periodic varied knot
states, i.e. at $t=0$ we have a knot on vortex-membrane-\textrm{A} that is
denoted by $\left \vert \sigma,\mathrm{A}\right \rangle ;$ at $t=\frac{\pi
}{\omega^{\ast}}$ we have a knot on vortex-membrane-\textrm{B} denoted by
$\left \vert \sigma,\mathrm{B}\right \rangle $. As a result, the leapfrogging
motion becomes a global winding along time direction, $\left \vert
t,\mathrm{A}\right \rangle ,$ $\left \vert t+\frac{\pi}{\omega^{\ast}%
},\mathrm{B}\right \rangle ,$ $\left \vert t+\frac{2\pi}{\omega^{\ast}%
},\mathrm{A}\right \rangle $, $\left \vert t+\frac{3\pi}{\omega^{\ast}%
},\mathrm{B}\right \rangle ,$ ... See the illustration of vortex-representation
of knot states for knot-crystal in Fig.2(c). After a time period $t=\frac{\pi
}{\omega^{\ast}},$ a knot state $\phi_{\mathrm{A}}\operatorname{mod}(2\pi
)\in(-\pi,0]$ turns into a knot state $\phi_{\mathrm{B}}\operatorname{mod}%
(2\pi)\in(-\pi,0].$ Thus, we use the following formulation to characterize the
leapfrogging process,
\begin{equation}
\psi_{\mathrm{A}}^{\dagger}\psi_{\mathrm{B}}^{\dagger}.
\end{equation}
After considering the energy from the leapfrogging process, a corresponding
term is given by%
\begin{equation}
2\hbar_{\mathrm{knot}}\omega^{\ast}\psi_{\mathrm{A}}^{\dagger}\psi
_{\mathrm{B}}^{\dagger}+h.c.
\end{equation}

From the global rotating motion denoted $e^{-i\omega_{0}t},$ the winding
states also change periodically. Because the contribution from global rotating
motion $e^{-i\omega_{0}t}$ is always canceled by shifting the chemical
potential, we don't consider its effect.

From above equation, in the limit $\left \vert \vec{k}\right \vert \rightarrow0$
we derive low energy effective Hamiltonian as
\begin{align}
\mathcal{H}_{\mathrm{3D}}  &  \simeq2aJ%
{\displaystyle \sum \limits_{k}}
\psi_{\mathrm{A,}k}^{\dagger}(\vec{\sigma}\cdot \vec{k})\psi_{\mathrm{A,}%
k}\nonumber \\
&  +2aJ%
{\displaystyle \sum \limits_{k}}
\psi_{\mathrm{B,}k}^{\dagger}(\vec{\sigma}\cdot \vec{k})\psi_{\mathrm{B,}%
k}\nonumber \\
&  +2\hbar_{\mathrm{knot}}\omega^{\ast}%
{\displaystyle \sum \nolimits_{k,\sigma}}
\psi_{\mathrm{A,\sigma,}k}^{\dagger}\psi_{\mathrm{B,\sigma,}k}^{\dagger}\\
&  =c_{\mathrm{eff}}\int \Psi^{\dagger}[\tau_{z}\otimes(\vec{\sigma}\cdot
\hat{k})]\Psi d^{3}x\nonumber \\
&  +m_{\mathrm{knot}}c_{\mathrm{eff}}^{2}\int \Psi^{\dagger}(\tau_{x}%
\otimes \vec{1})\Psi d^{3}x.
\end{align}
where%
\begin{equation}
\Psi(\mathbf{x})=\left(
\begin{array}
[c]{c}%
\psi_{\mathrm{A,}\uparrow}(t,\vec{x})\\
\psi_{\mathrm{B,}\uparrow}^{\ast}(t,\vec{x})\\
\psi_{\mathrm{A,}\downarrow}(t,\vec{x})\\
\psi_{\mathrm{B,}\downarrow}^{\ast}(t,\vec{x})
\end{array}
\right)  .
\end{equation}
We then re-write the effective Hamiltonian to be
\begin{equation}
\mathcal{H}_{\mathrm{3D}}=\int(\Psi^{\dagger}\hat{H}_{\mathrm{3D}}\Psi)d^{3}x
\end{equation}
and
\begin{equation}
\hat{H}_{\mathrm{3D}}=c_{\mathrm{eff}}\vec{\Gamma}\cdot \vec{p}_{\mathrm{knot}%
}+m_{\mathrm{knot}}c_{\mathrm{eff}}^{2}\Gamma^{5}%
\end{equation}
where
\begin{align}
\Gamma^{5}  &  =\tau^{x}\otimes \vec{1}\mathbf{,}\text{ }\Gamma^{1}=\tau
^{z}\otimes \sigma^{x},\\
\Gamma^{2}  &  =\tau^{z}\otimes \sigma^{y},\text{ }\Gamma^{3}=\tau^{z}%
\otimes \sigma^{z}.\nonumber
\end{align}
$\vec{p}=\hbar_{\mathrm{knot}}\vec{k}$ is the momentum operator.
$\Psi^{\dagger}=(\psi_{\mathrm{A},\uparrow}^{\ast},\psi_{\mathrm{B},\uparrow
},\psi_{\mathrm{A},\downarrow}^{\ast},\psi_{\mathrm{B},\downarrow})$ is the
annihilation operator of four-component fermions. $m_{\mathrm{knot}%
}c_{\mathrm{eff}}^{2}=2\hbar_{\mathrm{knot}}\omega^{\ast}$ plays role of the
mass of knots and $c_{\mathrm{eff}}=\frac{2a\cdot J}{\hbar_{\mathrm{knot}}}$
play the role of light speed where $a$ is a fixed length that denotes the half
pitch of the windings on the knot-crystal. In the following parts, we set
$\hbar_{\mathrm{knot}}=1$ and $c_{\mathrm{eff}}=1$.

Due to Lorentz invariance (see below discussion), the geometric space becomes
geometric space-time, i.e., $(x,y,z)\rightarrow(x,y,z,t)$. Here, we may
consider $\vec{\Gamma}$ and $\Gamma^{5}$ to be \emph{entanglement matrices}
along spatial and tempo direction in winding space-time, respectively. A
complete set of entanglement matrices $(\vec{\Gamma},$ $\Gamma^{5})$ is called
\emph{entanglement pattern}. The coordinate transformation along
x/y/z/t-direction is characterize by $e^{i\vec{\Gamma}\cdot \hat{k}\cdot \vec
{x}}$ and $e^{i\Gamma^{5}\cdot \hat{\omega}t}$, respectively. Now, the knot
becomes topological defect of 3+1D entanglement -- a knot is not only
anti-phase changing along arbitrary spatial direction $\vec{e}$ but also
becomes anti-phase changing along tempo direction (along tempo direction, a
knot switches a knot state $\left \vert \mathrm{A/B}\right \rangle $ to another
knot state $\left \vert \mathrm{B/A}\right \rangle $).

Finally, the low energy effective Lagrangian of 3D SOC knot-crystal is
obtained as
\begin{align}
\mathcal{L}_{\mathrm{3D}}  &  =i\Psi^{\dagger}\partial_{t}\Psi-\mathcal{H}%
_{\mathrm{3D}}\\
&  =\bar{\Psi}(i\gamma^{\mu}\hat{\partial}_{\mu}-m_{\mathrm{knot}}%
)\Psi \nonumber
\end{align}
where $\bar{\Psi}=\Psi^{\dagger}\gamma^{0},$ $\gamma^{\mu}$ are the reduced
Gamma matrices,
\begin{equation}
\gamma^{1}=\gamma^{0}\Gamma^{1},\text{ }\gamma^{2}=\gamma^{0}\Gamma^{2},\text{
}\gamma^{3}=\gamma^{0}\Gamma^{3},
\end{equation}
and
\begin{align}
\gamma^{0}  &  =\Gamma^{5}=\tau_{x}\otimes \vec{1},\\
\gamma^{5}  &  =i\gamma^{0}\gamma^{1}\gamma^{2}\gamma^{3}.\nonumber
\end{align}

In addition, we point out that there exists intrinsic relationship between the
knot states of sublattice-representation and the knot states of
vortex-representation%
\begin{equation}
\left(
\begin{array}
[c]{c}%
\left \vert \mathrm{A}\right \rangle \\
\left \vert \text{\textrm{B}}\right \rangle
\end{array}
\right)  =\mathcal{U}\left(
\begin{array}
[c]{c}%
\left \vert \mathrm{L}\right \rangle \\
\left \vert \text{\textrm{R}}\right \rangle
\end{array}
\right)
\end{equation}
where $\mathcal{U}=\exp[i\pi \left(
\begin{array}
[c]{cc}%
0 & -i\\
i & 0
\end{array}
\right)  ].$ From the sublattice-representation of knot states, the
knot-crystal becomes an object with staggered \textrm{R/L} zeroes along x/y/z
spatial directions and time direction; From the vortex-representation of knot
states, the knot-crystal becomes an object with global winding along x/y/z
spatial directions and time direction. See the illustration of knot states of
vortex-representation on a knot-crystal in Fig.2.

\subsubsection{Emergent Lorentz-invariance}

We discuss the emergent Lorentz-invariance for knot states on a knot-crystal.

Since the Fermi-velocity $c_{\mathrm{eff}}$ only depends on the microscopic
parameter $J$ and $a,$ we may regard $c_{\mathrm{eff}}$ to be "light-velocity"
and the invariance of light-velocity becomes an fundamental principle for the
knot physics. The Lagrangian for massive Dirac fermions indicates emergent
\textrm{SO(3,1)} Lorentz-invariance. The \textrm{SO(3,1)} Lorentz
transformations $S_{\mathrm{Lor}}$ is defined by
\begin{equation}
S_{\mathrm{Lor}}\gamma^{\mu}S_{\mathrm{Lor}}^{-1}=\gamma^{\prime \mu}%
\end{equation}
($\mu=0,1,2,3$) and%
\begin{equation}
S_{\mathrm{Lor}}\gamma^{5}S_{\mathrm{Lor}}^{-1}=\gamma^{5}.
\end{equation}

For a knot state with a global velocity $\vec{v},$ due to \textrm{SO(3,1)}
Lorentz-invariance, we can do Lorentz boosting on $(\vec{x},t)$ by considering
the velocity of a knot,
\begin{align}
t  &  \rightarrow t^{\prime}=\frac{t-\vec{x}\cdot \vec{v}}{\sqrt{1-\vec{v}^{2}%
}},\nonumber \\
\vec{x}  &  \rightarrow \vec{x}^{\prime}=\frac{\vec{x}-\vec{v}\cdot t}%
{\sqrt{1-\vec{v}^{2}}}.
\end{align}
We can do non-uniform Lorentz transformation $S_{\mathrm{Lor}}(\vec{x},t)$ on
knot states $\Psi(\vec{x},t).$ The \emph{inertial reference-frame} for quantum
states of the knot is defined under Lorentz boost, i.e.,
\begin{equation}
\Psi(\vec{x},t)\rightarrow \Psi^{\prime}(\vec{x}^{\prime},t^{\prime
})=S_{\mathrm{Lor}}\cdot \Psi(\vec{x}^{\prime},t^{\prime}).
\end{equation}
For a particle-like knot, a uniform wave-function of knot states $\psi(t)$ is
\begin{equation}
\psi(t)=\frac{1}{\sqrt{V}}e^{-i2\omega^{\ast}t}.
\end{equation}
Under Lorentz transformation with small velocity $\left \vert \vec
{v}\right \vert $, this wave-function $\psi(t)$ is transformed into
\begin{align}
\psi(t)  &  =\frac{1}{\sqrt{V}}e^{-i2\omega^{\ast}t}\nonumber \\
&  \rightarrow \psi^{\prime}=\frac{1}{\sqrt{V}}e^{-i2\omega^{\ast}t^{\prime}}\\
&  \simeq \frac{1}{\sqrt{V}}e^{-i2\omega^{\ast}t}\exp(-i(E_{\mathrm{knot}%
}t-\vec{p}_{\mathrm{knot}}\cdot \vec{x}))\nonumber
\end{align}
where $E_{\mathrm{knot}}\simeq \frac{\vec{p}_{\mathrm{knot}}^{2}}%
{2m_{\mathrm{knot}}},$ $\vec{p}_{\mathrm{knot}}\simeq \omega \vec{v}$ and
$m_{\mathrm{knot}}c^{2}=2\omega^{\ast}.$ As a result, we derive a new
distribution of knot-pieces by doing Lorentz transformation, that are
described by the plane-wave wave-function $\frac{1}{\sqrt{V}}e^{-i2\omega
^{\ast}t}\exp(-i(E_{\mathrm{knot}}t-\vec{p}_{\mathrm{knot}}\cdot \vec{x})).$
The new wave-function $\frac{1}{\sqrt{V}}\exp(-i(E_{\mathrm{knot}}t-\vec
{p}_{\mathrm{knot}}\cdot \vec{x}))$ comes from the Lorentz boosting
$S_{\mathrm{Lor}}$.

Noninertial system can be obtained by considering non-uniformly velocities,
i.e., $\vec{v}\rightarrow \Delta \vec{v}(\vec{x},t).$ According to the linear
dispersion for knots, we can do local Lorentz transformation on $(\vec{x},t)$
i.e.,
\begin{align}
t  &  \rightarrow t^{\prime}(\vec{x},t)=\frac{t-\vec{x}\cdot \Delta \vec{v}%
}{\sqrt{1-(\Delta \vec{v})^{2}}},\\
\vec{x}  &  \rightarrow \vec{x}^{\prime}(\vec{x},t)=\frac{\vec{x}-\Delta \vec
{v}\cdot t}{\sqrt{1-(\Delta \vec{v})^{2}}}.\nonumber
\end{align}
We can also do non-uniform Lorentz transformation $S_{\mathrm{Lor}}(\vec
{x},t)$ on knot states $\Psi(\vec{x},t),$ i.e.,
\begin{align}
\Psi(\vec{x},t)  &  \rightarrow \Psi^{\prime}(\vec{x}^{\prime}(\vec
{x},t),t^{\prime}(\vec{x},t))\\
&  =S_{\mathrm{Lor}}(\vec{x},t)\cdot \Psi(\vec{x},t)\nonumber
\end{align}
where the new wave-functions of all quantum states change following the
non-uniform Lorentz transformation $S_{\mathrm{Lor}}(\vec{x},t)$. It is
obvious that there exists intrinsic relationship between noninertial system
and curved space-time.

\subsection{Dirac model on winding space}

In this part, we show the effective Dirac model of knot states on winding space.

The coordinate measurement of zero-lattice on winding space is the winding
angles, $\vec{\Phi}$. Along a given direction $\vec{e}$, after shifting the
distance $a,$ the windng angle changes $\pi.$ The position is determined by
two kinds of values: $\vec{X}$ are integer numbers
\begin{equation}
\vec{X}=(X,Y,Z)=\frac{1}{\pi}\vec{\Phi}-\frac{1}{\pi}\vec{\Phi}%
\operatorname{mod}\pi
\end{equation}
and $\vec{\phi}$ denote internal winding angles
\begin{equation}
\vec{\phi}=(\phi_{x},\phi_{y},\phi_{z})=\vec{\Phi}\operatorname{mod}\pi
\end{equation}
with $\phi_{x},\phi_{y},\phi_{z}\in(0,\pi]$.

Therefore, on winding space, the effective Hamiltonian turns into
\begin{align}
\hat{H}_{\mathrm{3D}}  &  =\vec{\Gamma}\cdot \vec{p}_{\mathrm{knot}%
}+m_{\mathrm{knot}}\Gamma^{5}\nonumber \\
&  =\vec{\Gamma}\cdot \vec{p}_{X,\mathrm{knot}}+\vec{\Gamma}\cdot \vec{p}%
_{\phi,\mathrm{knot}}+m_{\mathrm{knot}}\Gamma^{5}%
\end{align}
where $\vec{p}_{X}=\frac{1}{a}i\frac{d}{d\vec{X}}$ and $\vec{p}_{\phi}%
=\frac{1}{a}i\frac{d}{d\vec{\phi}}$. Because of $\phi_{j}\in(0,\pi]$, quantum
number of $\vec{p}_{\phi}$ is angular momentum $\vec{L}_{\phi}$ and the energy
spectra are $\frac{1}{a}\left \vert \vec{L}_{\phi}\right \vert .$ If we focus on
the low energy physics $E\ll \frac{1}{a}$ (or $\vec{L}_{\phi}=0$), we may get
the low energy effective Hamiltonian as
\begin{equation}
\hat{H}_{\mathrm{3D}}\simeq \vec{\Gamma}\cdot \vec{p}_{X,\mathrm{knot}%
}+m_{\mathrm{knot}}\Gamma^{5}.
\end{equation}

We introduce \emph{3+1D winding space-time} by defining four coordinates on
winding space, $\Phi=(\vec{\Phi},\Phi_{t})$ where $\Phi_{t}$ is phase changing
under time evolution. For a fixed entanglement pattern $(\vec{\Gamma},$
$\Gamma^{5})$, the coordinate transformation along x/y/z/t-direction on
winding space-time is given by $e^{i\vec{\Gamma}\cdot \hat{\Phi}}$ and
$e^{i\Gamma^{5}\cdot \hat{\Phi}_{t}}$, respectively.

For low energy physics, the position in 3+1D winding space-time is 3+1D
zero-lattice of winding space-time labeled by four integer numbers,
$\mathbf{X}=(\vec{X},X_{0})$ where
\begin{align}
\vec{X}  &  =\frac{1}{\pi}\vec{\Phi}-\frac{1}{\pi}\vec{\Phi}\operatorname{mod}%
\pi,\text{ }\nonumber \\
X_{0}  &  =\frac{1}{\pi}\Phi_{t}-\frac{1}{\pi}\Phi_{t}\operatorname{mod}\pi.
\end{align}
The lattice constant of the winding space-time is always $\pi$ that will never
be changed. As a result, the winding space-time becomes an effective
\emph{quantized} space-time. Because of $x_{\mu}=a\cdot X_{\mu}$, the
effective action on 3+1D winding space-time becomes
\begin{equation}
\mathcal{S}_{\mathrm{3D}}\simeq(a)^{4}\sum_{X,Y,Z,X_{0}}\mathcal{L}%
_{\mathrm{3D}}%
\end{equation}
where%
\begin{equation}
\mathcal{L}_{\mathrm{3D}}=\bar{\Psi}[i\frac{1}{a}(\gamma^{\mu})\hat{\partial
}_{\mu}-m_{\mathrm{knot}}]\Psi.
\end{equation}

\section{Deformed zero-lattice as curved space-time}

In this section, we discuss the knot dynamics on smoothly deformed
knot-crystal (or deformed zero-lattice). We point out that to characterize the
entanglement evolution, the corresponding Biot-Savart mechanics for a knot on
smoothly deformed zero-lattice is mapped to that in quantum mechanics on a
curved space-time.

\subsection{Entanglement transformation}

Firstly, based on a uniform 3D knot-crystal (uniform entangled
vortex-membranes), we introduce the concept of "\emph{entanglement
transformation (ET)}".

Under global entanglement transformation, we have%
\begin{equation}
\Psi(\vec{x},t)\rightarrow \Psi^{\prime}(\vec{x},t)=\hat{U}_{\mathrm{ET}}%
(\vec{x},t)\cdot \Psi(\vec{x},t)
\end{equation}
where
\begin{equation}
\hat{U}_{\mathrm{ET}}(\vec{x},t)=e^{i\delta \vec{\Phi}\cdot \vec{\Gamma}}\cdot
e^{i\delta \Phi_{t}\cdot \Gamma^{5}}.
\end{equation}
Here, $\delta \vec{\Phi}$ and $\delta \Phi_{t}$ are constant winding angles
along spatial $\vec{\Phi}$-direction and that along tempo direction on
geometric space-time, respectively. The dispersion of the excitation changes
under global entanglement transformation.

In general, we may define (local) entanglement transformation, i.e.,
\begin{equation}
\hat{U}_{\mathrm{ET}}(\vec{x},t)=e^{i\delta \vec{\Phi}(\vec{x}.t)\cdot
\vec{\Gamma}}\cdot e^{i\delta \Phi_{t}(\vec{x}.t)\cdot \Gamma^{5}}%
\end{equation}
where $\delta \vec{\Phi}(\vec{x},t)$ and $\delta \Phi_{t}(\vec{x},t)$ are not
constant. We call a system with smoothly varied-($\delta \vec{\Phi}(\vec{x}%
,t)$, $\delta \Phi_{t}(\vec{x},t)$) \emph{deformed knot-crystal} and its
projected zero-lattice \emph{deformed (3+1D) zero-lattice}.

\subsection{Geometric description for deformed zero-lattice -- curved
space-time}

\begin{figure}[ptb]
\includegraphics[clip,width=0.53\textwidth]{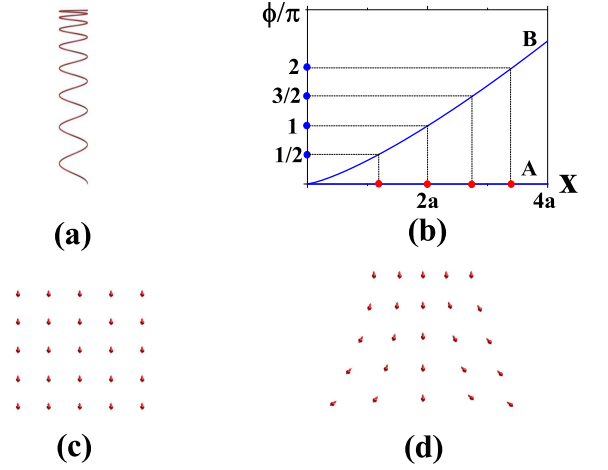}\caption{(a) An
illustration of deformed knot-crystal; (b) An illustration of smoothly
deformed relationship between winding angle $\Phi$ and spatial coordinate $x$.
The zero-lattice in winding space is still uniform; while the zero-lattice in
geometric space is deformed; (c) An illustration of a uniform 1+1D
zero-lattice in geometric space-time; (d) An illustration of a deformed 1+1D
zero-lattice in geometric space-time.}%
\end{figure}

For knots on a deformed zero-lattice, there exists an intrinsic correspondence
between an entanglement transformation $\hat{U}_{\mathrm{ET}}(\vec{x},t)$ and
a local coordinate transformation that becomes a fundamental principle for
emergent gravity theory in knot physics.

For zero-lattice, $\hat{U}_{\mathrm{ET}}(\vec{x},t)$ changes the winding
degrees of freedom that is denoted by the local coordination transformation,
i.e.,%
\begin{align}
\vec{\Phi}(\vec{x},t)  &  \implies \vec{\Phi}^{\prime}(\vec{x},t)=\vec{\Phi
}(\vec{x},t)+\delta \vec{\Phi}(\vec{x},t),\nonumber \\
\Phi_{t}(\vec{x},t)  &  \implies \Phi_{t}^{\prime}(\vec{x},t)=\Phi_{t}(\vec
{x},t)+\delta \Phi_{t}(\vec{x},t).
\end{align}
These equations also imply a curved space-time: the lattice constants of the
3+1D zero-lattice (the size of a lattice constant with $2\pi$ angle changing)
are not fixed to be $2a$, i.e.,
\[
2a\rightarrow2a_{\mathrm{eff}}(\vec{x},t)
\]
The distance between two nearest-neighbor "lattice sites" on the spatial/tempo
coordinate changes, i.e.,
\begin{align}
\Delta \vec{x}  &  =(\vec{x}+\vec{e}_{x})-\vec{x}=\vec{e}_{x},\\
\Delta \vec{x}^{\prime}  &  =(\vec{x}^{\prime}+\vec{e}_{x}^{\prime})-\vec
{x}^{\prime}=\vec{e}_{x}^{\prime}(\vec{x},t)\nonumber
\end{align}
and
\begin{align}
\Delta t  &  =(t+e_{0})-t=e_{0},\\
\Delta t^{\prime}  &  =(t^{\prime}+e_{0}^{\prime})-t^{\prime}=e_{0}^{\prime
}(\vec{x},t)\nonumber
\end{align}
where $e_{a}$ $(a=0,1,2,3)$ and $e_{a}^{\prime}(\vec{x},t)$ are the
unit-vectors of the original frame and the deformed frame, respectively. See
the illustration of a 1+1D deformed zero-lattice on winding space-time with a
non-uniform distribution of zeroes in Fig.3(d).

However, for deformed zero-lattice, the information of knots in projected
space is invariant: when the lattice-distance of zero-lattice changes
$a\rightarrow a_{\mathrm{eff}}(\vec{x},t)$, the size of the knots
correspondingly changes $a\rightarrow a_{\mathrm{eff}}(\vec{x},t)$. Therefore,
due to the invariance of a knot, the deformation of zero-lattice doesn't
change the formula of the low energy effective model for knots on winding
space-time. Because one may smoothly deform the zero-lattice and get the same
low energy effective model for knots on winding space-time, there exists
\emph{diffeomorphism invariance}, i.e.,
\begin{align}
&  \text{ Knot-invariance on winding space-time}\nonumber \\
&  \implies \text{Diffeomorphism invariance.}%
\end{align}
Therefore, from the view of mathematics, the physics on winding space-time is
never changed! The invariance of the effective model for knots on winding
space-time indicates the diffeomorphism invariance
\begin{equation}
\mathcal{S}_{\mathrm{zero-lattice}}\equiv(a)^{4}\sum_{X,Y,Z,X_{0}}\bar{\Psi
}[i\frac{1}{a}\gamma^{\mu}\hat{\partial}_{\mu}^{X}-m_{\mathrm{knot}}]\Psi.
\end{equation}

On the other hand, the condition of very smoothly entanglement transformation
guarantees a \emph{(local) Lorentz invariance} in long wave-length limit.
Under local Lorentz invariance, the knot-pieces of a given knot are determined
by local Lorentz transformations.

According to the local coordinate transformation, the deformed zero-lattice
becomes a curved space-time for the knots. In continuum limit $\Delta
k\ll(a)^{-1}$ and $\Delta \omega \ll \omega_{0}$, the diffeomorphism invariance
and (local) Lorentz invariance emerge together. E. Witten had made a strong
claim about emergent gravity, \textquotedblleft \textit{whatever we do, we are
not going to start with a conventional theory of non-gravitational fields in
Minkowski space-time and generate Einstein gravity as an emergent
phenomenon.}\textquotedblright \ He pointed out that gravity could be emergent
only if the notion on the space-time on which diffeomorphism invariance is
simultaneously emergent. For the emergent quantum gravity in knot physics,
diffeomorphism invariance and Lorentz invariance are simultaneously emergent.
In particular, the diffeomorphism invariance comes from information invariance
of knots on winding space-time -- when the lattice-distance of zero-lattice
changes, the size of the knots correspondingly changes.

To characterize the deformed 3+1D zero-lattice $\left(  \vec{x}^{\prime}%
(\vec{x},t),t^{\prime}(\vec{x},t)\right)  $, we introduce a geometric
description. In addition to the existence of a set of vierbein fields $e^{a}$,
the space metric is defined by $\eta_{ab}e_{\alpha}^{a}e_{\beta}^{b}%
=g_{\alpha \beta}$ where $\eta$ is the internal space metric tensor. The
geometry fields (vierbein fields $e^{a}(\vec{x},t)$ and spin connections
$\omega^{ab}(\vec{x},t)$) are determined by the non-uniform local coordinates
$\left(  \vec{x}^{\prime}(\vec{x},t),t^{\prime}(\vec{x},t)\right)  $.
Furthermore, one needs to introduce spin connections $\omega^{ab}(\vec{x},t)$
and the Riemann curvature 2-form as
\begin{align}
R_{b}^{a}  &  =d\omega_{b}^{a}+\omega_{c}^{a}\wedge \omega_{b}^{c}\\
&  =\frac{1}{2}R_{b\mu \nu}^{a}dx^{\mu}\wedge dx^{\nu},\nonumber
\end{align}
where $R_{b\mu \nu}^{a}\equiv e_{\alpha}^{a}e_{b}^{\beta}R_{\beta \mu \nu
}^{\alpha}$ are the components of the usual Riemann tensor projection on the
tangent space. The deformation of the zero-lattice is characterized by
\begin{equation}
R^{ab}=d\omega^{ab}+\omega^{ac}\wedge \omega^{cb}.
\end{equation}

So the low energy physics for knots on the deformed zero-lattice turns into
that for Dirac fermions on curved space-time
\begin{equation}
\mathcal{S}_{\mathrm{curved-ST}}=\int \sqrt{-g}\bar{\Psi}(e_{a}^{\mu}\gamma
^{a}(i\hat{\partial}_{\mu}+i\omega_{\mu})-m_{\mathrm{knot}})\Psi \text{ }d^{4}x
\end{equation}
where $\omega_{\mu}=(\omega_{\mu}^{0i}\gamma^{0i}/2,\omega_{\mu}^{ij}%
\gamma^{ij}/2)$ ($i,j=1,2,3$) and $\gamma^{ab}=-\frac{1}{4}[\gamma^{a}%
,\gamma^{b}]$ ($a,b=0,1,2,3$)\cite{23}. This model described by $\mathcal{S}%
_{\mathrm{curved-ST}}$ is invariant under local (non-compact) \textrm{SO(3,1)}
Lorentz transformation $S(\vec{x},t)=e^{\theta_{ab}(\vec{x},t)\gamma^{ab}}$
as
\begin{align}
\Psi(\vec{x},t)  &  \rightarrow \Psi^{\prime}(\vec{x},t)=S(\vec{x},t)\Psi
(\vec{x},t),\nonumber \\
\gamma^{\mu}  &  \rightarrow(\gamma^{\mu}(\vec{x},t))^{\prime}=S(\vec
{x},t)\gamma^{\mu}(S(\vec{x},t))^{-1},\nonumber \\
\omega_{\mu}  &  \rightarrow \omega_{\mu}^{\prime}(\vec{x},t)=S(\vec
{x},t)\omega_{\mu}(\vec{x},t)(S(\vec{x},t))^{-1}\nonumber \\
&  +S(\vec{x},t)\partial_{\mu}(S(\vec{x},t))^{-1}.
\end{align}
$\gamma^{5}$ is invariant under local \textrm{SO(3,1)} Lorentz symmetry as
\begin{align}
\gamma^{5}  &  \rightarrow(\gamma^{5})^{\prime}=S(\vec{x},t)\gamma^{5}%
(S(\vec{x},t))^{-1}\nonumber \\
&  =\gamma^{5}.
\end{align}
In general, an \textrm{SO(3,1)} Lorentz transformation $S(\vec{x},t)$ is a
combination of spin rotation transfromation $\hat{R}(\vec{x},t)=\hat
{R}_{\mathrm{spin}}(\vec{x},t)\cdot \hat{R}_{\mathrm{space}}(\vec{x},t)$ and
Lorentz boosting $S_{\mathrm{Lor}}(\vec{x},t)$.

In physics, under a Lorentz transformation, a distribution of knot-pieces
changes into another distribution of knot-pieces. For this reason, the
velocity $c_{\mathrm{eff}}$ and the total number of zeroes $N_{\mathrm{knot}}$
are invariant,
\begin{equation}
c_{\mathrm{eff}}\rightarrow c_{\mathrm{eff}}^{\prime}\equiv c_{\mathrm{eff}}%
\end{equation}
and
\begin{equation}
N_{\mathrm{knot}}\rightarrow N_{\mathrm{knot}}^{\prime}\equiv N_{\mathrm{knot}%
}.
\end{equation}

\subsection{Gauge description for deformed zero-lattice}

\subsubsection{Deformed entanglement matrices and deformed entanglement
pattern}

The deformation of the zero-lattice leads to deformation of entanglement
pattern, i.e.,
\begin{equation}
(\vec{\Gamma},\Gamma^{5})\rightarrow(\vec{\Gamma}^{\prime}(\mathbf{x}%
),(\Gamma^{5})^{\prime}(\mathbf{x}))
\end{equation}
where
\begin{align}
\vec{\Gamma}^{\prime}(\mathbf{x})  &  =\hat{U}_{\mathrm{ET}}(\mathbf{x}%
)\vec{\Gamma}\hat{U}_{\mathrm{ET}}(\mathbf{x})^{-1},\\
(\Gamma^{5})^{\prime}(\mathbf{x})  &  =\hat{U}_{\mathrm{ET}}(\vec{x}%
,t)\Gamma^{5}\hat{U}_{\mathrm{ET}}(\mathbf{x})^{-1}.\nonumber
\end{align}
$\mathbf{x}$ denotes the space-time position of a site of zero-lattice,
$(\vec{x},t)$. Each entanglement matrix becomes a unit \textrm{SO(4)}
vector-field on each lattice site. The deformed zero-lattice induced by local
entanglement transformation $\hat{U}_{\mathrm{ET}}(\mathbf{x})$ is
characterized by four \textrm{SO(4)} vector-fields (four entanglement
matrices) $(\vec{\Gamma}^{\prime}(\mathbf{x}),(\Gamma^{5})^{\prime}%
(\mathbf{x}))$. See the illustration of a 2D deformed zero-lattice in
Fig.(4)d, in which the arrows denote deformed entanglement matrix $(\Gamma
^{5})^{\prime}(\mathbf{x})$.

\subsubsection{Gauge description for deformed tempo entanglement matrix}

Firstly, we stduy the unit \textrm{SO(4)} vector-field of deformed tempo
entanglement matrix $(\Gamma^{5})^{\prime}(\mathbf{x}).$ To characterize
$(\Gamma^{5})^{\prime}(\mathbf{x}),$ the reduced Gamma matrices $\gamma^{\mu}$
is defined as
\begin{equation}
\gamma^{1}=\gamma^{0}\Gamma^{1},\text{ }\gamma^{2}=\gamma^{0}\Gamma^{2},\text{
}\gamma^{3}=\gamma^{0}\Gamma^{3},
\end{equation}
and
\begin{align}
\gamma^{0}  &  =\Gamma^{5}=\tau^{x}\otimes \vec{1},\\
\gamma^{5}  &  =i\gamma^{0}\gamma^{1}\gamma^{2}\gamma^{3}.\nonumber
\end{align}
Under this definition ($\gamma^{0}=\Gamma^{5}$), the effect of deformed
zero-lattice from spatial entanglement transformation $e^{i\Gamma^{1}%
\cdot \Delta \Phi_{x}}$, $e^{i\Gamma^{2}\cdot \Delta \Phi_{y}}$, $e^{i\Gamma
^{3}\cdot \Delta \Phi_{z}}$ can be studied due to
\begin{equation}
\Gamma^{5}\rightarrow(\Gamma^{5})^{\prime}(\mathbf{x})=\hat{U}_{\mathrm{ET}%
}^{x/y/z}(\vec{x},t)\Gamma^{5}\hat{U}_{\mathrm{ET}}^{x/y/z}(\mathbf{x}%
)^{-1}\neq \Gamma^{5}.
\end{equation}
However, the effect of deformed zero-lattice from tempo entanglement
transformation $e^{i\delta \Phi_{t}\cdot \Gamma^{5}}$ cannot be well defined due
to
\begin{equation}
\Gamma^{5}\rightarrow(\Gamma^{5})^{\prime}(\mathbf{x})=\hat{U}_{\mathrm{ET}%
}^{t}(\vec{x},t)\Gamma^{5}\hat{U}_{\mathrm{ET}}^{t}(\mathbf{x})^{-1}%
=\Gamma^{5}.
\end{equation}

We introduce an \textrm{SO(4)} transformation $\hat{U}(\vec{x},t)$ that is a
combination of spin rotation transfromation $\hat{R}(\mathbf{x})$ and spatial
entanglement transformation (entanglement transformation along
x/y/z-direction) $\hat{U}_{\mathrm{ET}}^{x/y/z}(\mathbf{x})=e^{i\delta
\vec{\Phi}(\mathbf{x})\cdot \vec{\Gamma}}$, i.e.,
\begin{equation}
\hat{U}(\mathbf{x})=\hat{R}(\mathbf{x})\oplus \hat{U}_{\mathrm{ET}}%
^{x/y/z}(\mathbf{x}).
\end{equation}
Here, $\oplus$ denotes operation combination. Under a non-uniform
\textrm{SO(4)} transformation $\hat{U}(\mathbf{x})$, we have\textrm{ }%
\begin{align}
\gamma^{0}  &  \rightarrow \hat{U}(\mathbf{x})\gamma^{0}(\hat{U}(\mathbf{x}%
))^{-1}\\
&  =(\gamma^{0}(\mathbf{x}))^{\prime}=%
{\displaystyle \sum \nolimits_{a}}
\gamma^{a}n^{a}(\mathbf{x})\nonumber
\end{align}
where $\mathbf{n}=(n^{1},n^{2},n^{3},\phi_{0}^{0})=(\vec{n},\phi_{0}^{0})$ is
a unit \textrm{SO(4)} vector-field. For the deformed zero-lattice, according
to $(\gamma^{0}(\mathbf{x}))^{\prime}\neq \gamma^{0},$ the entanglement matrix
$\Gamma^{5}=\gamma^{0}$ along tempo direction is varied, $\Gamma
^{5}\rightarrow(\Gamma^{5})^{\prime}(\mathbf{x})\neq \Gamma^{5}$.

In general, the \textrm{SO(4)} transformation is defined by $\hat
{U}(\mathbf{x})=e^{\Phi_{ab}(\mathbf{x})\gamma^{ab}}$ ($\gamma^{ab}=-\frac
{1}{4}[\gamma^{a},\gamma^{b}]$). Under the \textrm{SO(4)} transformation, we
have
\begin{align}
\gamma^{\mu}  &  \rightarrow(\gamma^{\mu}(\mathbf{x)})^{\prime}=\hat
{U}(\mathbf{x})\gamma^{\mu}(\hat{U}(\mathbf{x}))^{-1},\nonumber \\
A_{\mu}  &  \rightarrow A_{\mu}^{\prime}(\vec{x},t)=\hat{U}(\vec{x},t)A_{\mu
}(\mathbf{x})(\hat{U}(\mathbf{x}))^{-1}\nonumber \\
&  +\hat{U}(\mathbf{x})\partial_{\mu}(\hat{U}(\mathbf{x}))^{-1}.
\end{align}
In particular, $\gamma^{5}$ is invariant under the \textrm{SO(4)}
transformation as
\begin{align}
\gamma^{5}  &  \rightarrow(\gamma^{5})^{\prime}=\hat{U}(\mathbf{x})\gamma
^{5}(\hat{U}(\mathbf{x}))^{-1}\nonumber \\
&  =\gamma^{5}.
\end{align}
The correspondence between index of $\gamma^{a}$ and index of space-time
$x^{a}$ is
\begin{align}
\gamma^{1}  &  \Leftrightarrow x,\text{ }\gamma^{2}\Leftrightarrow y,\\
\gamma^{3}  &  \Leftrightarrow z,\text{ }\gamma^{0}\Leftrightarrow t.\nonumber
\end{align}
We denote this correspondence to be
\begin{equation}
(1,2,3,0)_{\mathrm{ET}}\Leftrightarrow(1,2,3,0)_{\mathrm{ST}}%
\end{equation}
where $(1,2,3,0)_{\mathrm{ET}}$ denotes the index order of $\gamma^{a}$ and
$(1,2,3,0)_{\mathrm{ST}}$ denotes the index order of space-time $x^{a}$.

As a result, we can introduce an auxiliary gauge field $A_{\mu}^{ab}%
(\mathbf{x})$ and use a gauge description to characterize the deformation of
the zero-lattice. The auxiliary gauge field $A_{\mu}^{ab}(\mathbf{x})$ is
written into two parts\cite{23}: \textrm{SO(3)} parts
\begin{equation}
A^{ij}(\mathbf{x})=\mathrm{tr}(\gamma^{ij}(\hat{U}(\mathbf{x}))d(\hat
{U}(\mathbf{x}))^{-1})
\end{equation}
and \textrm{SO(4)/SO(3)} parts
\begin{align}
A^{i0}(\mathbf{x})  &  =\mathrm{tr}(\gamma^{i0}\hat{U}(\mathbf{x}))d(\hat
{U}(\mathbf{x}))^{-1})\\
&  =\gamma^{0}d(\gamma^{i}(\mathbf{x}))^{\prime}=-\gamma^{i}d\left(
\gamma^{0}(\mathbf{x})\right)  ^{\prime}.\nonumber
\end{align}
The total field strength $\mathcal{F}^{ij}(\mathbf{x})$ of $i,j=1,2,3$
components can be divided into two parts
\begin{equation}
\mathcal{F}^{ij}(\mathbf{x})=F^{ij}+A^{i0}\wedge A^{j0}.
\end{equation}
According to pure gauge condition, we have Maurer-Cartan equation,
\begin{equation}
\mathcal{F}^{ij}(\mathbf{x})=F^{ij}+A^{i0}\wedge A^{j0}\equiv0
\end{equation}
or
\begin{align}
F^{ij}  &  =dA^{ij}+A^{ik}\wedge A^{kj}\\
&  \equiv-A^{i0}\wedge A^{j0}.\nonumber
\end{align}

Finally, we emphasize the equivalence between $\gamma^{0i}$ and $\Gamma^{i}$,
i.e., $\gamma^{0i}\Leftrightarrow \Gamma^{i}.$

\subsubsection{Gauge description for deformed spatial entanglement matrix}

Next, we stduy the unit \textrm{SO(4)} vector-field of deformed spatial
entanglement matrix $(\Gamma^{i})^{\prime}(\mathbf{x}).$ To characterize
$(\Gamma^{i})^{\prime}(\mathbf{x}),$ the reduced Gamma matrices $\gamma^{\mu}$
is defined as
\begin{equation}
\gamma^{1}=\gamma^{0}\Gamma^{j},\text{ }\gamma^{2}=\gamma^{0}\Gamma^{k},\text{
}\gamma^{3}=\gamma^{0}\Gamma^{5},
\end{equation}
and
\begin{align}
\gamma^{0}  &  =\Gamma^{i}=\tau^{z}\otimes \sigma^{i},\\
\gamma^{5}  &  =i\gamma^{0}\gamma^{1}\gamma^{2}\gamma^{3}.\nonumber
\end{align}
Here, $\Gamma^{i}$, $\Gamma^{j}$, and $\Gamma^{k}$ are three orthotropic
spatial entanglement matrices. Under this definition ($\gamma^{0}=\Gamma^{i}%
$), the effect of deformed zero-lattice from partial spatial/tempo
entanglement transformation $e^{i\Gamma^{j}\cdot \Delta \Phi_{j}}$,
$e^{i\Gamma^{k}\cdot \Delta \Phi_{k}}$, $e^{i\Gamma^{5}\cdot \Delta \Phi_{t}}$ can
be studied due to
\begin{equation}
\Gamma^{i}\rightarrow(\Gamma^{i})^{\prime}(\mathbf{x})=\hat{U}_{\mathrm{ET}%
}^{x_{j}/x_{k}/t}(\vec{x},t)\Gamma^{i}\hat{U}_{\mathrm{ET}}^{x_{j}/x_{k}%
/t}(\mathbf{x})^{-1}\neq \Gamma^{i}.
\end{equation}
However, the effect of deformed zero-lattice from spatial entanglement
transformation $e^{i\delta \Phi_{t}\cdot \Gamma^{5}}$ cannot be well defined due
to
\begin{equation}
\Gamma^{i}\rightarrow(\Gamma^{i})^{\prime}(\mathbf{x})=\hat{U}_{\mathrm{ET}%
}^{x_{i}}(\vec{x},t)\Gamma^{i}\hat{U}_{\mathrm{ET}}^{x_{i}}(\mathbf{x}%
)^{-1}=\Gamma^{i}.
\end{equation}

We use similar approach to introduce the gauge description. We can also define
the reduced Gamma matrices $\tilde{\gamma}^{\mu}$ as
\begin{equation}
\tilde{\gamma}^{1}=\tilde{\gamma}^{0}\Gamma^{2},\text{ }\tilde{\gamma}%
^{2}=\tilde{\gamma}^{0}\Gamma^{3},\text{ }\tilde{\gamma}^{3}=\tilde{\gamma
}^{0}\Gamma^{5},
\end{equation}
and
\begin{align}
\tilde{\gamma}^{0}  &  =\Gamma^{i}=\tau^{z}\otimes \sigma^{x},\\
\tilde{\gamma}^{5}  &  =i\tilde{\gamma}^{0}\tilde{\gamma}^{1}\tilde{\gamma
}^{2}\tilde{\gamma}^{3}.\nonumber
\end{align}
The correspondence between index of $\tilde{\gamma}^{a}$ and index of
space-time $x^{a}$ is
\begin{align}
\tilde{\gamma}^{1}  &  \Leftrightarrow y,\text{ }\tilde{\gamma}^{2}%
\Leftrightarrow z,\\
\tilde{\gamma}^{3}  &  \Leftrightarrow t,\text{ }\tilde{\gamma}^{0}%
\Leftrightarrow x.\nonumber
\end{align}
We denote this correspondence to be
\begin{equation}
(1,2,3,0)_{\mathrm{ET}}\Leftrightarrow(2,3,0,1)_{\mathrm{ST}}.
\end{equation}

Now, the \textrm{SO(4)} transformation $\tilde{U}(\vec{x},t)=e^{\Phi_{ab}%
(\vec{x},t)\tilde{\gamma}^{ab}}$ ($\tilde{\gamma}^{ab}=-\frac{1}{4}%
[\tilde{\gamma}^{a},\tilde{\gamma}^{b}]$) is not a combination of spin
rotation symmetry and entanglement transformation along x/y/z-direction.
However, for the case of $a$ or $b$ to be $0,$ $\tilde{U}(\vec{x}%
,t)=e^{\Phi_{a0}(\vec{x},t)\tilde{\gamma}^{a0}}$ denotes the entanglement
transformation along y/z/t-direction. The unit \textrm{SO(4)} vector-field on
each lattice site becomes\textrm{ }%
\begin{equation}
\tilde{U}(\mathbf{x})\tilde{\gamma}^{0}(\tilde{U}(\mathbf{x}))^{-1}%
=(\tilde{\gamma}^{0}(\mathbf{x}))^{\prime}=%
{\displaystyle \sum \nolimits_{a}}
\tilde{\gamma}^{a}\tilde{n}^{a}(\mathbf{x})
\end{equation}
where $\mathbf{\tilde{n}}=(\tilde{n}^{1},\tilde{n}^{2},\tilde{n}^{3}%
,\tilde{\phi}_{0}^{0})$ is a unit vector-field. The auxiliary gauge field
$\tilde{A}^{ab}(\mathbf{x})$ are defined by
\begin{equation}
\tilde{A}^{ab}(\mathbf{x})=\mathrm{tr}(\tilde{\gamma}^{ij}(\tilde
{U}(\mathbf{x}))d(\tilde{U}(\mathbf{x}))^{-1}).
\end{equation}
According to pure gauge condition, we also have the following Maurer-Cartan
equation,
\begin{equation}
\tilde{F}^{ij}=d\tilde{A}^{ij}+\tilde{A}^{ik}\wedge \tilde{A}^{kj}\equiv
-\tilde{A}^{i0}\wedge \tilde{A}^{j0}.
\end{equation}

Finally, we emphasize the equivalence between $\tilde{\gamma}^{0i}$ and
$\Gamma^{a}$, i.e., $\tilde{\gamma}^{01}\Leftrightarrow \Gamma^{2},$
$\tilde{\gamma}^{02}\Leftrightarrow \Gamma^{3},$ $\tilde{\gamma}^{03}%
\Leftrightarrow \Gamma^{5}$.

\subsubsection{Hidden \textrm{SO(4) }invariant for gauge description}

In addition, there exists a hidden global \textrm{SO(4) }invariant for
entanglement matrices along different directions in 3+1D (winding) space-time
$(\vec{\Gamma},\Gamma^{5})\rightarrow(\vec{\Gamma}^{\prime},(\Gamma
^{5})^{\prime})$. To show the hidden \textrm{SO(4) }invariant, we define the
reduced Gamma matrices $\tilde{\gamma}^{\mu}$ as
\begin{align}
\tilde{\gamma}^{1}  &  =\tilde{\gamma}^{0}\Gamma^{2},\text{ }\tilde{\gamma
}^{2}=\tilde{\gamma}^{0}\Gamma^{3},\text{ }\tilde{\gamma}^{3}=\tilde{\gamma
}^{0}\Gamma^{5},\\
\tilde{\gamma}^{0}  &  =\alpha \Gamma^{1}+\beta \Gamma^{2}+\gamma \Gamma
^{3}+\delta \Gamma^{5},\nonumber \\
\tilde{\gamma}^{5}  &  =i\tilde{\gamma}^{0}\tilde{\gamma}^{1}\tilde{\gamma
}^{2}\tilde{\gamma}^{3}\nonumber
\end{align}
with $\alpha^{2}+\beta^{2}+\gamma^{2}+\delta^{2}=1$. Here, $\alpha,$ $\beta,$
$\gamma,$ $\delta$ are constant.

Under this description, we can study the entanglement deformation along
orthotropic spatial/tempo directions to $x^{\prime}=\alpha x+\beta y+\gamma
z+\delta t$.

\subsection{Relationship between geometric description and gauge description
for deformed zero-lattice}

Due to the generalized spatial translation symmetry there exists an
\emph{intrinsic relationship} between gauge description for entanglement
deformation between two vortex-membranes and geometric description for global
coordinate transformation of the same deformed zero-lattice.

On the one hand, to characterize the changes of the positions of zeroes, we
must consider a curved space-time by using geometric description,
$\mathbf{x}=(\vec{x},t)\rightarrow \mathbf{x}^{\prime}=(\vec{x}^{\prime
},t^{\prime})$. On the other hand, we need to consider a varied vector-field
\begin{align}
(\gamma^{0}(\mathbf{x}))^{\prime}  &  =\hat{U}(\mathbf{x})\gamma^{0}(\hat
{U}(\mathbf{x}))^{-1}\\
&  =%
{\displaystyle \sum \nolimits_{a}}
\gamma^{a}n^{a}(\mathbf{x})\nonumber
\end{align}
by using gauge description. There exists intrinsic relationship between the
geometry fields $e^{a}(\mathbf{x})$ ($a=1,2,3,0$) and the auxiliary gauge
fields $A^{a0}(\mathbf{x})$.

For a non-uniform zero-lattice, we have
\begin{align}
\vec{\Phi}(\vec{x},t)  &  \implies \vec{\Phi}^{\prime}(\vec{x},t)=\vec{\Phi
}(\vec{x},t)+\delta \vec{\Phi}(\vec{x},t),\nonumber \\
\Phi_{t}(\vec{x},t)  &  \implies \Phi_{t}^{\prime}(\vec{x},t)=\Phi_{t}(\vec
{x},t)+\delta \Phi_{t}(\vec{x},t).
\end{align}
On deformed zero-lattice, the "lattice distances" become dynamic vector
fields. We define the vierbein fields $e^{a}(\mathbf{x})$ that are supposed to
transform homogeneously under the local symmetry, and to behave as ordinary
vectors under local entanglement transformation along $x^{a}$-direction,
\begin{equation}
e^{a}(\mathbf{x})=dx^{a}(\mathbf{x})=\frac{a}{\pi}d\Phi^{a}(\mathbf{x}).
\end{equation}

For the smoothly deformed vector-fields $n^{i}(\mathbf{x})\ll1$, within the
representation of $\Gamma^{5}=\gamma^{0}$\ we have
\begin{align}
\frac{d\Phi^{i}(\mathbf{x})}{2\pi}  &  =n^{i}(\mathbf{x})=\mathrm{tr}%
[\gamma^{0}d\gamma^{i}(\mathbf{x})]\nonumber \\
&  =A^{i0}(\mathbf{x}),\text{ }i=1,2,3.
\end{align}
Thus, the relationship between $e^{i}(\mathbf{x})$ and $A^{i0}(\mathbf{x})$ is
obtained as
\begin{equation}
e^{i}(\mathbf{x})\equiv(2a)A^{i0}(\mathbf{x}).
\end{equation}
According to this relationship, the changing of entanglement of the
vortex-membranes curves the 3D space.

On the other hand, within the representation of $\Gamma^{i}=\tilde{\gamma}%
^{0}$\ we have%
\begin{align}
\frac{d\Phi^{a}(\mathbf{x})}{2\pi}  &  =\tilde{n}^{a}(\mathbf{x}%
)=\mathrm{tr}[\tilde{\gamma}^{0}d\tilde{\gamma}^{a}(\mathbf{x})]\nonumber \\
&  =\tilde{A}^{i0}(\mathbf{x}),\text{ }i=j,k,0,
\end{align}
and
\begin{equation}
e^{0}(\mathbf{x})=dt(\mathbf{x})=\frac{a}{\pi}d\Phi_{t}(\mathbf{x}%
)=(2a)\tilde{A}^{30}(\mathbf{x}).
\end{equation}
According to this relationship, the changing of entanglement of the
vortex-membranes curves the 4D space-time.

In addition, we point out that for different representation of reduced Gamma
matrix, there exists intrinsic relationships between the gauge fields
$A(\mathbf{x})$ and $\tilde{A}(\mathbf{x}).$ After considering these
relationships, we have a complete description of the deformed zero-lattice on
the geometric space-time,

\section{Emergent gravity}

Gravity is a natural phenomenon by which all objects attract one another
including galaxies, stars, human-being and even elementary particles. Hundreds
of years ago, Newton discovered the inverse-square law of universal
gravitation, $F=\frac{GMm}{r^{2}}$ where $G$ is the Newton constant, $r$ is
the distance, and $M$ and $m$ are the masses for two objects. One hundred
years ago, the establishment of general relativity by Einstein is a milestone
to learn the underlying physics of gravity that provides a unified description
of gravity as a geometric property of space-time. From Einstein's equations
$R_{\mu \nu}-\frac{1}{2}Rg_{\mu \nu}=8\pi GT_{\mu \nu},$ the gravitational force
is really an effect of curved space-time. Here $R_{\mu \nu}$ is the 2nd rank
Ricci tensor, $R$ is the curvature scalar, $g_{\mu \nu}$ is the metric tensor,
and $T_{\mu \nu}$ is the energy-momentum tensor of matter.

In this section, we point out that there exists emergent gravity for knots on zero-lattice.

\subsection{Knots as topological defects}

\subsubsection{Knot as \textrm{SO(4)/SO(3)} topological defect in 3+1D
space-time}

A knot corresponds to an elementary object of a knot-crystal; A knot-crystal
can be regarded as composite system of multi-knot. For example, for 1D knot,
people divide the knot-crystal into $N$ identical pieces, each of which is
just a knot.

From point view of \emph{information}, each knot corresponds to a zero between
two vortex-membranes along the given direction. For a knot, there must exist a
zero point, at which $\xi_{\mathrm{A}}(x)$ is equal to $\xi_{\mathrm{B}}(x)$.
The position of the zero is determined by a local solution of the
zero-equation, $F_{\theta}(x)=0$ or $\xi_{\mathrm{A},\theta}(x)=\xi
_{\mathrm{B},\theta}(x).$

From point view of \emph{geometry}, a knot (an anti-knot) removes (or adds) a
projected zero of zero-lattice that corresponds to removes (or adds) half of
"lattice unit" on the zero-lattice according to
\begin{equation}
\Delta x_{i}=\pm a_{\mathrm{eff}}(\vec{x},t)\simeq \pm a.
\end{equation}
As a result, a knot looks like a special type of edge dislocation on 3+1D
zero-lattice. The zero-lattice is deformed and becomes mismatch with an
additional knot.

From point view of \emph{entanglement}, a knot becomes topological defect of
3+1D winding spacet-time: along $x$-direction, knot is anti-phase changing
denoted by $e^{i\Gamma^{1}\cdot \Delta \Phi_{x}}$, $\Delta \Phi_{x}=\pi;$ along
$y$-direction, knot is anti-phase changing denoted by $e^{i\Gamma^{2}%
\cdot \Delta \Phi_{y}}$, $\Delta \Phi_{y}=\pi;$ along $z$-direction, knot is
anti-phase changing denoted by $e^{i\Gamma^{3}\cdot \Delta \Phi_{z}}$,
$\Delta \Phi_{z}=\pi;$ along $t$-direction, knot is anti-phase changing denoted
by $e^{i\Gamma^{5}\cdot \Delta \Phi_{t}},$ $\Delta \Phi_{t}=\pi.$ Fig.4(a) and
Fig.4(b) show an illustration a 1D knot.

In mathematics, to generate a knot at $(x_{0},y_{0},z_{0},t_{0})$, we do
global topological operation on the knot-crystal, i.e.,
\begin{equation}
e^{i\Gamma^{1}\cdot \Delta \Phi_{x}(\mathbf{x})}\left \vert 0\right \rangle
\end{equation}
with $\Delta \Phi_{x}=0,$ $x<x_{0}$ and $\Delta \Phi_{x}=\pi,$ $x\geq x_{0};$
\begin{equation}
e^{i\Gamma^{2}\cdot \Delta \Phi_{y}(\mathbf{x})}\left \vert 0\right \rangle
\end{equation}
with $\Delta \Phi_{y}=0,$ $y<y_{0}$ and $\Delta \Phi_{y}=\pi,$ $y\geq y_{0};$
\begin{equation}
e^{i\Gamma^{3}\cdot \Delta \Phi_{z}(\mathbf{x})}\left \vert 0\right \rangle
\end{equation}
with $\Delta \Phi_{z}=0,$ $z<z_{0}$ and $\Delta \Phi_{z}=\pi,$ $z\geq x_{0};$
\begin{equation}
e^{i\Gamma^{5}\cdot \Delta \Phi_{t}(\mathbf{x})}\left \vert 0\right \rangle
\end{equation}
with $\Delta \Phi_{t}=0,$ $t<t_{0}$ and $\Delta \Phi_{t}=\pi,$ $t\geq t_{0}.$ As
a result, due to the rotation symmetry in 3+1D space-time, a knot becomes
\textrm{SO(4)/SO(3)} topologcal defect. Along arbitrary direction, the local
entanglement matrices around a knot at center are switched on the tangentia sub-space-time.

\subsubsection{Knot as \textrm{SO(3)/SO(2)} magnetic monopole in 3D space}

To characterize the topological property of a knot on the 3+1D zero-lattice,
we use gauge description. We firstly study the tempo entanglement deformation
and define $\Gamma^{5}=\gamma_{0}.$ Under this gauge description, we can only
study the effect of a knot on three spatial zero-lattice.\begin{figure}[ptb]
\includegraphics[clip,width=0.53\textwidth]{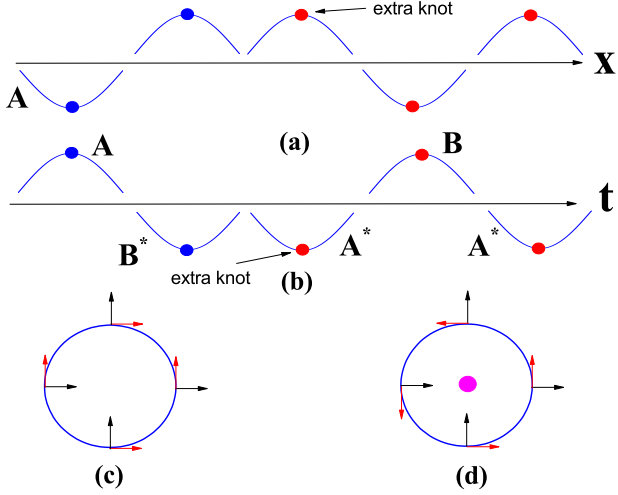}\caption{(a) An
illustration of the effect of an extra knot on a 1D knot-crystal along spatial
direction; (b) An illustration of the effect of an extra knot on a 1D
knot-crystal along tempo direction. Here \textrm{A}$^{\ast}$/\textrm{B}%
$^{\ast}$ denotes conjugate representation of vortex-line-\textrm{A/B}; (c)
The entanglement pattern for a uniform knot-crystal. The arrows denote the
directions of entanglement matrices; (d) The entanglement pattern for a
knot-crystal with an extra knot at center. The purple spot denotes the knot.
The red arrows that denote local tangential entanglement matrices have
vortex-like configuration on 2D projected space.}%
\end{figure}

When there exists a knot, the periodic boundary condition of knot states along
arbitrary direction is changed into anti-periodic boundary condition,
\begin{equation}
\Delta \Phi_{x}=\pi,\text{ }\Delta \Phi_{y}=\pi,\text{ }\Delta \Phi_{z}=\pi.
\end{equation}
Consequently, along given direction (for example $x$-direction), the local
entanglement matrices on the tangential sub-space are switched by
$e^{i\Gamma^{1}\cdot \Delta \Phi_{x}}$ ($\Delta \Phi_{x}=\pi$). Along
$x$-direction, in the limit of $x\rightarrow-\infty$, we have the local
entanglement matrices on the tangential sub-space as $\Gamma^{2}$ and
$\Gamma^{3}$; in the limit of $x\rightarrow \infty$, we have the local
entanglement matrices on the tangential sub-space as $e^{i\Gamma^{1}%
\cdot \Delta \Phi_{x}}(\Gamma^{2})e^{-i\Gamma^{1}\cdot \Delta \Phi_{x}}%
=-\Gamma^{2}$ and $e^{i\Gamma^{1}\cdot \Delta \Phi_{x}}(\Gamma^{3}%
)e^{-i\Gamma^{1}\cdot \Delta \Phi_{x}}=-\Gamma^{3}$.

Because we have similar result along $x^{i}$-direction for the system with an
extra knot, the system has generalized spatial rotation symmetry. Due to the
generalized spatial rotation symmetry, when moving around the knot, the local
tangential entanglement matrices (we may use indices $j$, $k$ to denote the
sub space) must rotate synchronously. See the red arrows that denote local
tangential entanglement matrices in Fig.4(c) and Fig.4(d). In Fig.4(d), local
tangential entanglement matrices induced by an extra (unified) knot shows
vortex-like topological configuration in projected 2D space (for example, x-y
plane). As a result, local tangential entanglement matrices induced by an
extra knot can be exactly mapped onto that of an orientable sphere with fixed chirality.

To characterize the topological property of 3+1D zero-lattice with an extra
(unified) knot, we apply gauge description and write down the following
constraint%
\begin{equation}%
{\displaystyle \iiint}
\rho_{F}dV=\frac{1}{4\pi}%
{\displaystyle \iint}
\epsilon_{jk}\epsilon_{ijk}F_{jk}^{jk}\cdot dS_{i} \label{monopole}%
\end{equation}
where
\begin{align}
F^{ij}  &  =dA^{ij}+A^{ik}\wedge A^{kj}\\
&  \equiv-A^{i0}\wedge A^{j0}\nonumber
\end{align}
and $\rho_{F}=\sqrt{-g}\psi^{\dagger}\psi.$ The upper indices of $F_{jk}^{jk}$
label the local entanglement matrices on the tangential sub-space and the
lower indices of $F_{jk}^{jk}$ denote the spatial direction. The non-zero
Gaussian integrate $\frac{1}{4\pi}%
{\displaystyle \iint}
\epsilon_{jk}\epsilon_{ijk}F_{jk}^{jk}\cdot dS_{i}$ just indicates the local
entanglement matrices on the tangential sub-space $A^{i0}\wedge A^{j0}$ to be
the local frame of an orientable sphere with fixed chirality.

As a result, the entanglement pattern with an extra 3D knot is topologically
deformed and the 3D knot becomes \textrm{SO(3)/SO(2)} \emph{magnetic monopole.
}From the point view of gauge description, a knot traps a "magnetic charge" of
the auxiliary gauge field, i.e.,
\begin{equation}
N_{F}=\int \sqrt{-g}\Psi^{\dagger}\Psi d^{3}x=q_{m}%
\end{equation}
where $q_{m}=\frac{1}{4\pi}%
{\displaystyle \iint}
\epsilon_{jk}\epsilon_{ijk}F_{jk}^{jk}\cdot dS_{i}$ is the "magnetic" charge
of auxiliary gauge field $A^{jk}$. For single knot $N_{F}=1$, the "magnetic"
charge is $q_{m}=1$.

\subsubsection{Knot as \textrm{SO(3)/SO(2)} magnetic monopole in 2+1D
space-time}

Next, we study the spatial entanglement deformation and define $\Gamma
^{i}=\tilde{\gamma}_{0}.$ Under this gauge description, we can only study the
effect of a knot on 2D spatial zero-lattice and 1D tempo zero-lattice.

In the 2+1D space-time, a knot also leads to $\pi$-phase changing,%
\begin{equation}
\Delta \Phi_{i}=\pi,\text{ }\Delta \Phi_{j}=\pi,\text{ }\Delta \Phi_{t}=\pi.
\end{equation}
Due to the spatial-tempo rotation symmetry, the knot also becomes
\textrm{SO(3)/SO(2)} magnetic monopole and traps a "magnetic charge" of the
auxiliary gauge field $\tilde{A}^{jk}$, i.e.,
\begin{equation}
N_{F}=\int \sqrt{-g}\Psi^{\dagger}\Psi d^{3}x=\tilde{q}_{m}%
\end{equation}
where $\tilde{q}_{m}$ is the "magnetic" charge of auxiliary gauge field
$\tilde{A}^{ij}$. Remember that the correspondence between index of
$\tilde{\gamma}^{i}$ and index of space-time $x^{i}$ is $\tilde{\gamma}%
^{1}\Leftrightarrow y,$ $\tilde{\gamma}^{2}\Leftrightarrow z,$ $\tilde{\gamma
}^{3}\Leftrightarrow t.$

To characterize the induced magnetic charge, we write down another constraint%
\begin{equation}%
{\displaystyle \iiint}
\rho_{F}dV=\frac{1}{4\pi}%
{\displaystyle \iint}
\epsilon_{ij}\epsilon_{ijk}\tilde{F}_{jk}^{ij}\cdot dS_{i} \label{monopole1}%
\end{equation}
where
\begin{align}
\tilde{F}^{ij}  &  =d\tilde{A}^{ij}+\tilde{A}^{ij}\wedge \tilde{A}^{ij}\\
&  \equiv-\tilde{A}^{i0}\wedge \tilde{A}^{j0}.\nonumber
\end{align}
The upper indices of $\tilde{F}^{ij}=d\tilde{F}^{ij}+\tilde{F}^{ik}%
\wedge \tilde{F}^{kj}$ denote the local entanglement matrices on the tangential
sub-space-time and the lower indices of $\tilde{F}_{jk}^{ij}$ denote the
spatial direction. Therefore, according to above equation, the 2+1D
zero-lattice is globally deformed by an extra knot.

In general, due to the hidden \textrm{SO(4) }invariant, for other gauge
descriptions $\tilde{\gamma}^{0}=\alpha \Gamma^{1}+\beta \Gamma^{2}+\gamma
\Gamma^{3}+\delta \Gamma^{5}$, a knot also play the role of
\textrm{SO(3)/SO(2)} magnetic monopole and traps a "magnetic charge" of the
corresponding auxiliary gauge field.

\subsection{Einstein-Hilbert action as topological mutual BF term for knots}

It is known that for a given gauge description, a knot is an
\textrm{SO(3)/SO(2)} magnetic monopole and traps a "magnetic charge" of the
corresponding auxiliary gauge field. For a complete basis of entanglement
pattern, we must use four orthotropic \textrm{SO(4)} rotors $((\Gamma
^{1})^{\prime}(\mathbf{x}),(\Gamma^{2})^{\prime}(\mathbf{x}),(\Gamma
^{3})^{\prime}(\mathbf{x}),(\Gamma^{5})^{\prime}(\mathbf{x}))$ and four
different gauge descriptions to characterize the deformation of a knot (an
\textrm{SO(4)/SO(3)} topological defect) on a 3+1D zero-lattice.

Firstly, we use Lagrangian approach to characterize the deformation of a knot
(an \textrm{SO(3)/SO(2)} topological defect) on a 3D spatial zero-lattice,
$N_{F}=q_{m}$. The topological constraint in Eq.(\ref{monopole}) can be
re-written into
\begin{equation}
\frac{i}{4}\text{\textrm{tr}}\sqrt{-g}\bar{\Psi}\gamma^{i}(\gamma^{0i}%
/2)\Psi=\epsilon_{jk}\epsilon_{ijk}\frac{1}{4\pi}\hat{D}_{i}F_{jk}^{jk}%
\end{equation}
or
\begin{equation}
\frac{i}{4}\text{\textrm{tr}}\sqrt{-g}\bar{\Psi}\varpi_{0}^{0i}\gamma
^{i}(\gamma^{0i}/2)\Psi=i\epsilon_{0ijk}\epsilon_{0ijk}\varpi_{0}^{0i}\frac
{1}{4\pi}\hat{D}_{i}F_{jk}^{jk} \label{12}%
\end{equation}
where $\hat{D}_{i}=i\hat{\partial}_{i}+i\omega_{i}$ is covariant derivative in
3+1D space-time. $\varpi^{0i}$ is a field that plays the role of Lagrangian
multiplier. The upper index $i$ of $\varpi^{0i}$ denotes the local radial
entanglement matrix around a knot, along which the entanglement matrix doesn't
change. Thus, we use the dual field $\varpi^{0i}$ to enforce the topological
constraint in Eq.(\ref{monopole}). That is, to denote the upper index of
$F^{jk}$ that is\ the local tangential entanglement matrices,\ we set
antisymmetric property of upper index of $\varpi^{0i}$ and that of $F^{jk}$.
Because $\varpi^{0i}$ and $\omega^{0i}$ have the same \textrm{SO(3,1)}
generator $(\gamma^{0i}/2)$, due to \textrm{SO(3,1)} Lorentz invariance we can
do Lorentz transformation and absorb the dual field $\varpi^{0i}$ into
$\omega^{0i}$, i.e., $\omega^{0i}\rightarrow(\omega^{0i})^{\prime}=\omega
^{0i}+\varpi^{0i}$. As a result, the dual field $\varpi^{0i}$ is replaced by
$\omega^{0i}$.

In the path-integral formulation, to enforce such topological constraint, we
may add a topological mutual BF term $S_{\mathrm{MBF}}$ in the action that is
\begin{align}
S_{\mathrm{MBF1}}  &  =-\frac{1}{4\pi}\int \epsilon_{0ijk}\, \epsilon
_{0\nu \lambda \kappa}\,R_{0\nu}^{0i}F_{\lambda \kappa}^{jk}\text{ }d^{4}x\\
&  =-\frac{1}{4\pi}\int \epsilon_{0ijk}\,R^{0i}\wedge F^{jk}\nonumber
\end{align}
where
\begin{equation}
R^{0i}=d\omega^{0i}+\omega^{0j}\wedge \omega^{ji}.
\end{equation}
From $F^{jk}\equiv-A^{j0}\wedge A^{k0}$ and $e^{i}\wedge e^{j}=(2a)^{2}%
A^{j0}\wedge A^{k0}.$ The induced topological mutual BF term $S_{\mathrm{MBF1}%
}$ is linear in the conventional strength in $R^{0i}$ and $F^{jk}$. This term
is becomes
\begin{equation}
S_{\mathrm{MBF1}}=\frac{1}{4\pi(2a)^{2}}\int \epsilon_{0ijk}R^{0i}\wedge
e^{j}\wedge e^{k}.
\end{equation}

Next, we use Lagrangian approach to characterize the deformation of a knot (an
\textrm{SO(3)/SO(2)} topological defect) on 2+1D space-time, $N_{F}=\tilde
{q}_{m}$. Using the similar approach, we derive another topological mutual BF
term $S_{\mathrm{MBF2}}$ in the action that is
\begin{align}
S_{\mathrm{MBF2}}  &  =-\frac{1}{4\pi}\int \epsilon_{0ijk}\, \epsilon
_{0\nu \lambda \kappa}\, \tilde{R}_{0\nu}^{0i}\tilde{F}_{\lambda \kappa}%
^{jk}\text{ }d^{4}x\\
&  =-\frac{1}{4\pi}\int \epsilon_{0ijk}\, \tilde{R}^{0i}\wedge \tilde{F}%
^{jk}\nonumber
\end{align}
where $\tilde{R}^{0i}=d\tilde{\omega}^{0i}+\tilde{\omega}^{0j}\wedge
\tilde{\omega}^{ji}$. From $\tilde{F}^{k0}\equiv-\tilde{A}^{kj}\wedge \tilde
{A}^{j0}$ and $\tilde{e}^{i}\wedge \tilde{e}^{j}=(2a)^{2}\tilde{A}^{j0}%
\wedge \tilde{A}^{k0},$ this term becomes
\begin{equation}
S_{\mathrm{MBF2}}=\frac{1}{4\pi(2a)^{2}}\int \epsilon_{ijk0}\tilde{R}%
^{0i}\wedge \tilde{e}^{j}\wedge \tilde{e}^{k}.
\end{equation}
The upper index of $\tilde{R}^{0i}$ denotes entanglement transformation along
given direction in winding space-time. We unify the index order of space-time
into $(1,2,3,0)_{\mathrm{ST}}$ and reorganize the upper index. The topological
mutual BF term becomes $\frac{1}{4\pi(2a)^{2}}\int \epsilon_{ijk0}R^{ij}\wedge
e^{k}\wedge e^{0}.$ In Ref.\cite{mm,mm1,mm2,mm4}, a topological description of
Einstein-Hilbert action is proposed by S. W. MacDowell and F. Mansouri. The
topological mutual BF term proposed in this paper is quite different from the
MacDowell-Mansouri action.

According to the diffeomorphism invariance of winding space-time, there exists
symmetry between the entanglement transformation along different directions.
Therefore, with the help of a complete set of definition of reduced Gamma
matrices $\gamma^{\mu},$ there exist other topological mutual BF terms
$S_{\mathrm{MBF},i}.$ For the total topological mutual BF term
$S_{\mathrm{MBF}}=%
{\displaystyle \sum \limits_{i}}
S_{\mathrm{MBF},i}$ that characterizes the deformation of a knot (an
\textrm{SO(4)/SO(3)} topological defect) on a 3+1D zero-lattice, the upper
index of the topological mutual BF term $R^{ij}\wedge e^{k}\wedge e^{l}$ must
be symmetric, i.e., $i,j,k,l=1,2,3,0$.

By considering the \textrm{SO(3,1)} Lorentz invariance, the topological mutual
BF term $S_{\mathrm{MBF}}$ turns into the Einstein-Hilbert action
$S_{\mathrm{EH}}$ as
\begin{align}
S_{\mathrm{MBF}}  &  =S_{\mathrm{EH}}=\frac{1}{16\pi(a)^{2}}\int
\epsilon_{ijkl}R^{ij}\wedge e^{k}\wedge e^{l}\\
&  =\frac{1}{16\pi G}\int \sqrt{-g}Rd^{4}x\nonumber
\end{align}
where $G$ is the induced Newton constant which is $G=a^{2}.$ The role of the
Planck length is played by $l_{p}=a$, that is the "lattice" constant on the
3+1D zero-lattice.

Finally, from above discussion, we derived an effective theory of knots on
deformed zero-lattice in continuum limit as%
\begin{align}
S  &  =\mathcal{S}_{\mathrm{zero-lattice}}+S_{\mathrm{EH}}\\
&  =\int \sqrt{-g(x)}\bar{\Psi}(e_{a}^{\mu}\gamma^{a}\hat{D}_{\mu
}-m_{\mathrm{knot}})\Psi \text{ }d^{4}x\nonumber \\
&  +\frac{1}{16\pi G}\int \sqrt{-g}R\text{ }d^{4}x\nonumber
\end{align}
where $\hat{D}_{\mu}=i\hat{\partial}_{\mu}+i\omega_{\mu}$. The variation of
the action $S$ via the metric $\delta g_{\mu \nu}$ gives the Einstein's
equations
\begin{equation}
R_{\mu \nu}-\frac{1}{2}Rg_{\mu \nu}=8\pi GT_{\mu \nu}.
\end{equation}
As a result, in continuum limit a knot-crystal becomes a space-time background
like smooth manifold with emergent Lorentz invariance, of which the effective
gravity theory turns into \emph{topological} field theory.

For emergent gravity in knot physics, an important property is topological
interplay between zero-lattice and knots. From the Einstein-Hilbert action, we
found that the key property is duality between Riemann curvature $R^{ij}$ and
strength of auxiliary gauge field $F^{kl}$: \emph{the deformation of
entanglement pattern leads to the deformation of space-time}.

In addition, there exist a natural energy cutoff $\hbar \omega_{0}$ and a
natural length cutoff $a$. In high energy limit ($\Delta \omega \sim \omega_{0}$)
or in short range limit ($\Delta x\sim a$), without well-defined 3+1D
zero-lattice, there doesn't exist emergent gravity at all.

\section{Discussion and conclusion}

In this paper, we pointed out that owing to the existence of local Lorentz
invariance and diffeomorphism invariance there exists emergent gravity for
knots on 3+1D zero-lattice. In knot physics, the emergent gravity theory is
really a topological theory of entanglement deformation. For emergent gravity
theory in knot physics, a topological interplay between 3+1D zero-lattice and
the knots appears: on the one hand, the deformation of the 3+1D zero-lattice
leads to the changes of knot-motions that can be denoted by curved space-time;
on the other hand, the knots trapping topological defects deform the 3+1D
zero-lattice that indicates matter may curve space-time. The Einstein-Hilbert
action $S_{\mathrm{EH}}$ becomes a topological mutual BF term $S_{\mathrm{MBF}%
}$ that exactly reproduces the low energy physics of the general relativity.
In table.1, we emphasize the relationship between modern physics and knot physics.

\begin{widetext}
\begin{table*}[t]%
\begin{tabular}
[c]{|c|cccc|}\hline Modern physics& Knot Physics & & &\\
\hline Matter & Knot: a topological defect of 3+1 D zero-lattice && &
\\ \hline Motion& Changing of the distribution of knot-pieces & & &
\\ \hline Mass& Angular frequency for leapfrogging motion & & &
\\ \hline Inertial reference system& A knot under Lorentz boosting & & &
\\ \hline Coordinate translation& Entanglement transformation& & &
\\ \hline Space-time& 3+1D zero-lattice of projected entangled vortex-membranes & & &
\\ \hline Curved space-time& Deformed 3+1D zero-lattice & & &
\\ \hline Gravity& Topological interplay between 3+1D zero-lattice and knots & &
&\\ \hline
\end{tabular}
\caption{The relationship between modern physics and knot physics}
\end{table*}
\end{widetext}

In addition, this work would help researchers to understand the mystery in
gravity. In modern physics, matter and space-time are two \emph{different}
fundamental objects and matter may move in (flat or curved) space-time. In
knot physics, the most important physics idea for gravity is the unification
of matter and space-time, i.e.,
\begin{equation}
\text{Matter (knots)}\iff \text{Space-time (zero-lattice).}%
\end{equation}
One can see that matter (knots) and space-time (zero-lattice) together with
motion of matter are \emph{unified} into a simple phenomenon -- entangled
vortex-membranes and matter (knots) curves space-time (3+1D zero-lattice) via
a \emph{topological} way.

\acknowledgments This work is supported by NSFC Grant No. 11674026.


\begin{thebibliography}{99}
\bibitem {23}In this paper, we use $I=x,y,z$, $i,j=1,2,3$ and, $a,b=1,2,3,0.$
\end{thebibliography}
\end{document}